%% file: paper-outline.tex
\documentclass[pra,twocolumn]{revtex4-2}

\usepackage[en-US,useregional,showdow]{datetime2}
\DTMlangsetup[en-US]{ord=raise}

\usepackage{standalone}

\usepackage{amsthm}
\theoremstyle{definition}

\theoremstyle{plain}

\usepackage[utf8]{inputenc}
\usepackage[T1]{fontenc}
\usepackage{graphicx}
\usepackage{wrapfig}
\usepackage{rotating}
\usepackage[normalem]{ulem}
\usepackage{amsmath}
\usepackage{amssymb}
\usepackage{capt-of}
\usepackage{hyperref}
\usepackage{amsthm}
\usepackage[dvipsnames]{xcolor}
\usepackage{physics}
\usepackage{slashed}
\usepackage{siunitx}
\usepackage{mathtools}
\usepackage{enumitem}
\usepackage[capitalize]{cleveref}

\usepackage[caption=false,justification=raggedright]{subfig}

\usepackage[margin=2cm]{geometry}
\usepackage{figures/lib/mjsketch}

\newcommand{\ZIIS}{\mathbb{Z}_2^{(S)}}
\newcommand{\SITK}{\(\mathsf{SI1000}\)}

\begin{document}
\title{Practical Fault-Tolerant Gates by Majorana Fermion Motion}

\author{Yuri D. Lensky}
\affiliation{Google Quantum AI}
\author{Bryce Kobrin}
\affiliation{Google Quantum AI}
\author{Kostyantyn Kechedzhi}
\affiliation{Google Quantum AI}
\author{Igor Aleiner}
\affiliation{Google Quantum AI}


\begin{abstract}
  Quantum error correction protocols protect against local errors by storing
  logical information non-locally. This poses a challenge: how to design
  efficient logical gates on the non-local ``hidden'' logical information,
  and how to implement these gates using the local physical operations.
  We develop a general description of planar Pauli stabilizer codes and
  protocols for logical operations in terms of point-like particles called
  Majorana fermions. Information is stored in the pairwise fermion parities
  of spatially separated Majorana fermions. The description in terms of
  Majorana fermions captures not only large distance asymptotics, but also
  all scales down to the lattice constants. We exploit this locality to
  densely pack logical information in spacetime. The simplest application is
  to a static case: dense memory. More importantly, we implement
  fault-tolerant Majorana motion and leverage this primitive to design
  braiding-based logical gates. This approach reduces space overhead of
  logical operations resulting in an improved logical error rate given fixed
  number of physical qubits. We illustrate a practical use of our approach by
  designing and benchmarking of 2-qubit Clifford gates. We find numerically
  that our protocol outperforms lattice surgery in this setting for near-term
  error rates and realistic device constraints. More generally, introduction
  of compact motion of Majorana fermions as an efficient computational
  primitive opens a promising new route for the design of low overhead error
  correction protocols.
\end{abstract}

\maketitle

\section{Introduction}

In the practical design of fault-tolerant computations, one encounters two
complementary tasks: how to encode information efficiently, and how to
operate on the encoded information.  QEC codes with low hardware overhead are
realized with low-weight stabilizer measurements compiled into elementary
one- and two-qubit gate and measurement operations. To facilitate this task,
Kitaev proposed an asymptotic approach to fault-tolerant
computation\cite{kitaevFaulttolerantQuantumComputation2003}: information is
stored in fusion outcomes of non-Abelian anyons, while logical operations are
realized as motion of the anyons, implementing braids, fusion, and pair
creation. In this work, we focus on designing paths for the anyon motion for
specific physical device, and compilation of the motion into
hardware-specific circuits. Optimization of such paths and circuits is
challenging. Our approach to these challenges is based on a theoretical
framework developed in \cite{LENSKY2023169286}. It enables the construction
and characterization of 2D Pauli codes solely from the locations of
flux-bound Majorana fermions, which realize non-Abelian anyons in this
setting. Therefore in our codes, motion, pair creation, and fusion of
Majorana fermions realize fault-tolerant logical operations.

A remarkable feature of our formalism is that asymptotic properties are exact
on the lattice \cite{LENSKY2023169286}. The Majorana fermions whose fusion
outcomes form the logical Hilbert space are explicitly identified and
localized to single physical qubits. Importantly, code distances can be
determined with precision up to the lattice constant. Additionally, the
Majorana fermions can be individually manipulated. By combining these
ingredients we are able to design compact braids which implement the full
2-qubit Clifford gate set fault-tolerantly.

We will show that the resulting codes, motions, and overall protocols can be
efficiently implemented on hardware designed for the standard surface code,
scalable realizations of which were recently demonstrated in two different
hardware platforms, superconducting qubits~\cite{acharyaQuantumErrorCorrection2025} and neutral atoms~\cite{Bluvstein2023}. Another useful property of this construction is that
errors can be decoded by the matching-based decoders used for the surface
code~\cite{dennisTopologicalQuantumMemory2002}. This is because single-qubit Pauli errors create (at most) a pair of
Abelian anyons~\footnote{The \(e,m,\varepsilon\) anyons of the surface code~\cite{dennisTopologicalQuantumMemory2002}.}
in both cases. In this sense our approach is backwards-compatible with the
surface code, both at the level of hardware and software requirements, as
well as interpretation of existing protocols.

In particular, the standard lattice surgery gate protocols for the surface
code implement Pauli measurements in the logical space, and in our formalism
these operations can be understood as fusion of Majorana fermions. This is a
special case of measuring the fusion outcome that does not require Majorana
fermion motion to be fault tolerant. A natural next step would be to attempt
to reduce the overhead necessary for such measurement-based computation. This
could be done using the techniques we describe in this work, but the same
analysis shows that there is an even lower overhead approach. Specifically,
we use our formalism to include braiding and more generally Majorana fermion
motion as an additional generator of logical operations. In this approach we
exploit locality of Majorana fermions to increase encoding rate by a finite
factor compared to the surface code, and we exploit topological properties of
anyon braiding to realize compact logical Clifford gates. A combination of
braiding and tightly packing the Majorana fermions results in improved
scaling of logical error rate (of memory and Clifford gates) compared to
lattice surgery.~\footnote{The construction of
  compact braids contradicts the conclusions of
  Ref.~\cite{HorsmanFowler_2012}. However, this reference discussed braiding
  of defects rather than point-like Majorana fermions. The former apparently requires
  significantly larger overhead.}

To keep discussions concrete, we will focus on a very specific problem:
implementing the full 2-qubit Clifford gateset. We present a specification of
this protocol in terms of Majorana paths for devices with square-grid
connectivity, and discuss circuit construction and scaling of the protocol
for general error models. We provide a concrete implementation for an
architecture with realistic \cite{acharyaQuantumErrorCorrection2025}
constraints, and numerically benchmark this implementation with the widely
used \SITK{} error model. We compare fidelities with lattice surgery on
devices of various sizes. We find that for a wide range of parameters
braiding-based protocols demonstrate higher fidelity Clifford gates compared
to their alternative implementation through lattice surgery.

Before describing our results in detail, we comment on the relationship of
our work to prior results in the literature. It has long been understood
(first identified in~\cite{kitaevAnyonsExactlySolved2006}) that there are
certain ``twist'' or symmetry defects that can be embedded in the surface
code that act as non-Abelian Ising
anyons~\textcite{bombinTopologicalOrderTwist2010,
  kitaevModelsGappedBoundaries2012a}. This has motivated efforts to exploit
the properties of these defects for
computation~\cite{barkeshliTheoryDefectsAbelian2013,youProjectiveNonAbelianStatistics2012a,hastings2015reducedspacetimetimecosts,benhemouNonAbelianStatisticsMixedboundary2021,zhengDemonstratingNonAbelianStatistics2015,brownPokingHolesCutting2017a}. However,
a concrete general scheme for manipulating and using these anyons, suitable
both for macroscopic protocol-design level and efficient circuit
construction, has not been put forward. Moreover, concrete protocols for
fault tolerant 2-qubit unitary gates suggested so far have been less qubit
efficient than lattice surgery. In this work we address both problems, via a
microscopic theory of the basic degrees of freedom for more general 2D Pauli
codes (generalizing the non-Abelian anyons beyond a symmetry defect to the
fundamental particle).

\section{Majoranas as a macroscopic degree of freedom}
\label{sec:majoranas-as-macro-dof}

We start by describing the degrees of freedom which are relevant for error
corrected memory and gates in 2D Pauli codes. We call these particles
Majorana fermions (they will be formally introduced in the next section). We
describe our protocols in terms of their locations in the plane. Memory
design consists of laying out these particles in 2D. Logical Clifford gates,
measurement, and initialization are implemented in terms of braids, fusions,
and pair creations of the Majorana fermions, and their design consists of
laying out \emph{paths} for these particles in spacetime.

Beyond a convenient visualization of protocols, this approach provides a
rough heuristic for logical fidelty by bounding code distances. The bounds
are simply computed in terms of a graph (to be specified in more detail in
the next section) on which the Majorana fermions are placed. The code
distance on disk topology~\footnote{In other topologies, one has to account
  for additional paths of both types that wrap handles, holes, or go between
  boundaries.}  is computed as the minimum of two lengths;
\(\ell_c = \min \{ \ell_W, \ell_H \}\).  The length \(\ell_W\) is the minimal
graph distance between particles~\footnote{As we will see in
  \cref{sec:majorana-fermions-as-micro-dof}, counted as the number of
  vertices along the path.}, which we will refer to as Wilson
lines\footnote{In \cite{LENSKY2023169286}, it was important to distinguish
  between 2 types of Wilson lines, while we will only be concerned with 1
  type. When we refer to Wilson lines in this work, we are referring to what
  are called ``augmented'' Wilson lines in~\cite{LENSKY2023169286}. The
  ``unaugmented'' Wilson lines are the conventional parallel transport
  operators, whereas ``augmented'' Wilson lines are gauge-invariant (since
  they are either loops or terminated by matter fields).}. To compute
\(\ell_H\), construct the shortest path \(\gamma_H\) through the dual lattice
which (i) crosses an even number of edges of the original lattice, (ii) is
either mounted on the boundaries or is a loop, and (iii) separates the
particles into two nonempty groups. Then \(\ell_H\) is (bounded below by) the
smallest set of \emph{vertices} \(V\) such that every \emph{edge} of
\(\gamma_H\) touches a vertex of \(V\). We will refer to such paths as 't
Hooft lines.

Although the \(O(1)\) distances depend on the details of the graph, in
certain cases it is useful to instead consider a fixed graph. This
effectively fixes a metric for the two sorts of paths. This is often a good
starting point for the design of a protocol. This is the case for the primary
examples of this paper, where all the essential large-distance features are
captured by metrics induced by a square grid. The induced metrics are a
Manhattan metric for computing \(\ell_W\), and a \(\pi/4\) rotated Manhattan
metric \(\ell_H\).

\begin{figure}[htbp!]

  \subfloat[Surface code single qubit memory.\label{fig:surface-code-ir-mjs}]{
    \input{figures/graphical-notation-sc-mjs}
  }
  \subfloat[Two qubit memory\label{fig:2q-ir-mjs}]{
    \input{figures/6-mj-mjs}
  } \\
  \subfloat[Dense three qubit memory.\label{fig:dense-3q-ir-mjs}]{
    \input{figures/3q-mj-mjs}
  }
  \caption{Several memories written in terms of Majorana fermions. The
    vertical dimension in all the figures is \(d\) (in the metric described
    in the main text), and is \(2d\) (\(d\)) for
    \cref{fig:2q-ir-mjs,fig:dense-3q-ir-mjs}
    (\cref{fig:surface-code-ir-mjs}). The wavy lines are Wilson lines (see
    main text). A choice for which Pauli operator they correspond in the
    logical encoding is shown in
    \cref{fig:surface-code-ir-mjs,fig:2q-ir-mjs}. All blue and yellow Wilson
    lines have the same length.}
  \label{fig:small-memories}
\end{figure}
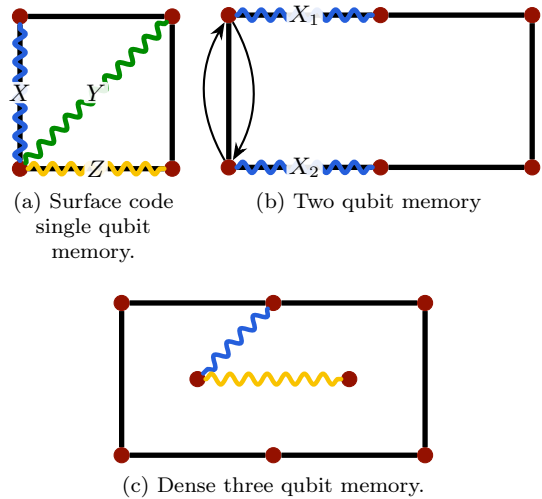

We demonstrate how the macroscopic properties of memories arise from these
metrics by analyzing three examples. The number of logical qubits in these
examples turns out to be \(n_{\sigma} / 2 - 1\), where \(n_{\sigma}\) is the
number of particles. The first example, \cref{fig:surface-code-ir-mjs}, is
actually the standard surface code encoding a single logical qubit. The
second example, \cref{fig:2q-ir-mjs}, has an extra logical qubit, and it is
easy to check that it has the same code distance by using the metrics. The
3rd example, \cref{fig:dense-3q-ir-mjs}, is more intriguing, and illustrates
some key properties of these metrics that will be essential for our gate
protocols. It has the same geometric dimensions as the second picture (at
fixed code distance \(d\)), but encodes 3 qubits. Note that if we had the
Euclidean metric, the distance would be reduced to \(d/\sqrt{2}\). In fact,
this is a maximally dense packing at distance \(d\) according to the metrics
above, the fact apparently overlooked in the previous literature. For example, both Wilson lines shown are length \(d\). It is denser
than the first example by a factor of \(3/2\). The density can be further
improved for encodings of more qubits by combining the dense packing of
Majoranas with manifold topology (holes or handles). A complimentary paper Ref.~\cite{low2026denserplanarsurfacecode} considered application of dense memory to reduce overhead of quantum simulation of chemistry.

Particle motion is a sufficient primitive to implement the full logical
Clifford gateset fault-tolerantly. We focus on braiding, which generates the
Clifford group on 2 logical qubits in the setup of \cref{fig:2q-ir-mjs}. We
aim to optimize logical gate fidelty. This optimization is fundamentally
device-specific. To simplify it, we focus on several heirarchies of distances
(which are correlated to the logical fidelity). The most detailed is a
``circuit-level'' distance. In this section we instead discuss a simpler
upper bound on this distance that is constructed as follows. We represent the
particle motion as a sequence of snapshots of particle locations in time.  We
only allow the particles to move a finite distance between snapshots. The
bound is the smallest distance amongst the snapshots. We would like to have
as dense an encoding in each snapshot as possible, while leaving enough room
to perform logical operations. Our approach is to first optimize the upper
bound, then attempt to saturate it.

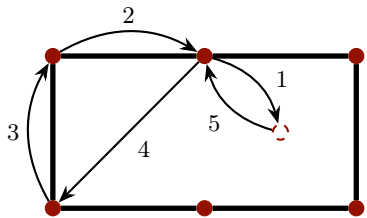
\begin{figure}[htbp]
  \centering
  \input{figures/braiding-mj-picture}
  \caption{Path of a distance-preserving braid in the space of 2 qubit memory
    \cref{fig:2q-ir-mjs}, i.e. 2 surface code patches. The steps can be
    broken up in order as shown, but in fact they can also be partially
    parallelized while preserving the distance.}
  \label{fig:braid-outline}
\end{figure}

We demonstrate a solution of this optimization problem in the case that
illustrates all the key difficulties: implementing the full Clifford gateset
on 2 logical qubits in presence of boundaries. Specifically, we choose paths
for the motions that implement the gates while maximizing code distance, and
provide concrete circuits that implement these motions. As an example, say
that we wish to exchange the two particles on the left shown
\cref{fig:2q-ir-mjs}, while keeping them as far as possible. Even if we had
the Euclidean metric, we might choose the path shown
\cref{fig:braid-outline}. The distance would be reduced from
\(d \to d / \sqrt{2}\). The Manhattan metrics for the two types of paths
induced by the square grid suggests that in fact the braid is
distance-preserving, and therefore optimal (as far as code distance). To
generate the full Clifford group, one additionally needs either the swap of a
horizontal pair, or a cyclic permutation of 4 Majoranas. There is a simpler
bulk braid (which would be the primary operation in a larger system), however
additional braids are needed to generate the logical Clifford group. Note
that in the absence of boundaries it is sufficient to increase the spacing
between densely packed particles by \(O(1)\) to allow sufficient room for
distance preserving braiding. This suggests one possible route to
architecture with reduced physical qubit count. In the remainder of this
paper, we elaborate on the microscopic theory and demonstrate concrete
circuits that implement this braid achieving the distance \(d\).

We can also discuss the scaling of the procedure in time, still as a function
of code distance. It is clear that the leading scaling is \(\propto d\), with
the prefactor determined by how quickly the Majorana fermions can be
moved. Define \(v_{\sigma}\) so that in time \(d / v_{\sigma}\) we can move a
distance \(d\) (while preserving the code distance). Then the prefactor can
be made at most \((2 + 1) / v_{\sigma}\) by parallelizing the motion. The
optimization of \(v_{\sigma}\) and details of parallelization is
architecture- and error-model specific, and we leave the detailed analysis of
this to future work.

Finally, measurement and initialization are implemented by pair annihilation
and creation, respectively. The fundamental primitive is still Majorana
fermion motion. We note that the effect of distance on fidelity of processes
involving pair creation or annihilation is more complicated, but in the most
important practical cases the relevant distance is the minimal one between
the particles other than the pair being created or annihilated.

\section{Majorana fermions as microscopic degrees of freedom}
\label{sec:majorana-fermions-as-micro-dof}

The purpose of this Section is to microscopically justify the classical heuristics described in Sec.~\ref{sec:majoranas-as-macro-dof}. The results are based on Ref.~\cite{LENSKY2023169286} and summary of minimum necessary details is included in Appendix. The main result is the prescription how to compute microscopic lengths of Wilson and 't Hooft lines and explain their significance for arbitrary planar lattices. As we will see it is useful to consider lattices that differ from the square lattice by local rearrangement of bonds. 

We start by describing the fundamental degrees of freedom: Majorana fermions, denoted \(\sigma\). They are fundamental in the sense that all other excitations discussed in the context of toric code  \(e, m\) and
\(\varepsilon\) are emergent from $\sigma$. 

Majorana fermions have the fusion rule
\begin{equation}
  \label{eq:mj-fusion-rule}
  \sigma \times \sigma = 1 + \varepsilon,
\end{equation}
where the fermion \(\varepsilon\) is essentially defined by this
rule. Information can be stored in the fusion outcomes of pairs of Majorana
fermions, meaning that a pair Majorana fermions act in a two dimensional Hilbert space that can be labeled by presence or absence of the fermion $\varepsilon$.

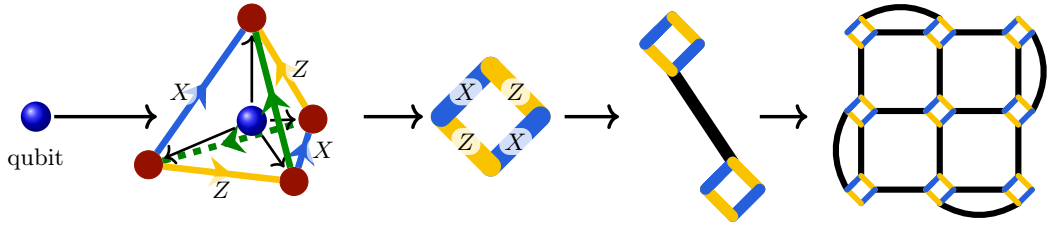
\begin{figure*}[t]
  \centering
  \input{figures/gauge-theory-outline}
  \caption{Outline of the steps of the Majorana embedding, described in more
    detail in \cite{LENSKY2023169286}. First, a set of arrowed tetrahedra for
    \(n\) qubits is actually a diagrammatic representation of a specific
    embedding of their Hilbert space into that of \(4n\) Majorana
    fermions. Majorana bilinears associated to a single qubit, shown with
    colors here and elsewhere, are images of Pauli operators under this
    embedding. This requires a choice of ordering on the bilinears,
    represented as arrows in the picture. The choice of such arrows gives
    rise to a background gauge field we call \(\mathbb{Z}_2^{(K)}\). The
    systematic procedure for choosing a gauge and physical consequences of
    this gauge field are discussed in \cite{LENSKY2023169286}, and for
    clarity we suppress arrows in following pictures. Disjoint Majorana
    bilinears commute, and the constraint associated to the embedding given
    by the tetrahedra is that bilinears of the same color represent the same
    operator. In the Majorana language, this is a gauge constraint fixing the
    fermion parity at each site. More general bilinears (which may include
    bilinears on the same qubit) are shown as black edges. A collection of
    disjoint black edges, or Majorana dimers, is a Majorana graph. Such
    bilinears are in general not invariant under the parity constraint. They
    are parallel transport operators for a gauge field we call
    \(\ZIIS\)\cite{LENSKY2023169286}. The rightmost picture shows the
    Majorana graph for a distance 3 surface code. Products of black edge
    operators around loops measure \(\ZIIS\) fluxes and correspond to the
    standard stabilizers of this code.}
  \label{fig:mj-gauge-embedding}
\end{figure*}

Our immediate goal is to reduce the description of a system of \(n\) qubits
to the \(n_{\sigma} \ll n\) Majorana fermions of
\cref{sec:majoranas-as-macro-dof}. This is accomplished by viewing the entire
system as \(4n\) Majorana fermions, subject to two kinds of local (gauge)
constraints. Each is associated to a gauge field, which we label \(\mathbb{Z}_2^{(K)}\)
and \(\ZIIS\). The Majorana fermions are non-Abelian
because they are charged and bound to flux of \(\mathbb{Z}_2^{(K)}\). Errors
are detected as fluxes of \(\ZIIS\).

It is simplest to describe these constraints diagramatically,
and for this purpose we introduce a Majorana graph. We follow the sequence
shown in \cref{fig:mj-gauge-embedding} (see \cite{LENSKY2023169286} for more
details on the formal structure). First, we assign 4 Majorana fermions to
each qubit, and the first gauge constraint is to fix the parity sector to be even  for
each such group. Majorana bilinears on a single qubit (shown with colors in
\cref{fig:mj-gauge-embedding} and elsewhere) are gauge-invariant and satisfy
a Pauli algebra on each qubit.

The Majorana graph is a collection of dimers between these Majorana fermions,
shown as black edges in \cref{fig:mj-gauge-embedding} and elsewhere. They are parallel transport operators for the field \(\ZIIS\) (so are
  not gauge invariant if they go between different qubits).
Therefore, products of dimers around loops
measure the fluxes of \(\ZIIS\). Since dimers do not share Majorana fermions
they commute individually, and therefore all loops commute with each
other. They act as Pauli strings in the gauge invariant space and form a
basis of a Pauli stabilizer code.  In the no-flux sector for \(\ZIIS\), the
gauge-invariant outcomes of fusion of the Majorana dimers are fixed to \(1\),
see~\cite{LENSKY2023169286}. Although there are \(4n\) Majorana fermions,
only undimerized Majorana fermions, $n_\sigma$, can encode logical
qubits. Therefore on a disk topology~\footnote{If we embed in a general
  manifold \(M\) and only take topologically trivial loops as stabilizers,
  the number of logical qubits scales as \(\sim n_{\sigma} / 2 - \chi(M)\)
  with \(O(1)\) corrections (related to threaded Majorana ``fluxes'')
  discussed in Ref.~\cite{LENSKY2023169286}.}, the code has
\(\max \left\{ \frac{n_{\sigma}}{2} - 1 , 0 \right\}\) logical qubits. This
is just the even parity sector of the fusion outcomes of pairings of the
\(n_{\sigma}\) undimerized Majorana fermions. For the rest of the paper, when
we refer to Majorana fermions, we are referring to the undimerized ones
unless explicitly indicated otherwise. For brevity, we will state results
only for the disk topologies.

The code protects information by separating the Majorana fermions. To
quantify this statement, we proceed to describe the action of Pauli errors
and how to bound the code distance, as well as derive the metrics from
\cref{sec:majoranas-as-macro-dof}. Detectable errors simply create different
combinations of \(\varepsilon\) fusion outcomes for dimerized pairs,
detectable as fluxes of the \(\ZIIS\) gauge field. We first discuss the
situation away from any boundary or \(\sigma\).
The code is then locally the toric code, which is often discussed in terms of
the fusion outcome \(\varepsilon\), and the emergent \(e,m\) excitations. We
describe it using the fundamental degrees of freedom of the Majorana code,
the Majorana fermions and their fusion outcomes \(1,\varepsilon\).

\begin{figure}[t]
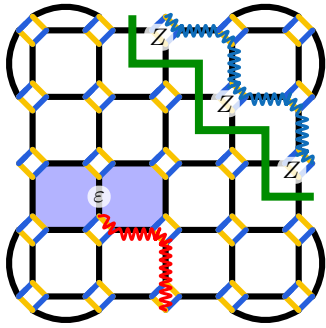

  \centering
  \include{figures/code-distance-tHoof-line}
  \caption{An example code where code distance is determined by the 't Hooft
    line shown in green, that is shorter than the Wilson line shown in
    blue. Pauli $Z$ label errors that correspond to the 't Hooft
    line. Following the rules of Sec.~\ref{sec:majoranas-as-macro-dof} we
    determine that $\ell_H =3$ whereas $\ell_W =5$. We also show an example
    Wilson line in red, which has one end on an undimerized Majorana fermion
    and another on a dimerized one. The procedure for writing Wilson lines as
    Pauli strings is detailed in \cite{LENSKY2023169286}.}
  \label{fig:tHooft-line-boundary}
\end{figure}

Changing the fusion
outcome associated to a single dimerized pair from \(1\) to \(\varepsilon\)
flips the contribution of the dimer between them to the \(\ZIIS\) stabilizer
loops.
Therefore, a single \(\varepsilon\) is locally represented by a pair of
edge-adjacent \(\ZIIS\) fluxes.  Fermion parity is a gauge constraint, so
(away from any \(\sigma\)) gauge-invariant operators can only create pairs of
\(\varepsilon\).
At a boundary, the only difference is that the fusion outcome of a boundary
dimer is detected by only one flux. This is in principle a complete
description of the action of Pauli errors on Majorana graph states, away from
any \(\sigma\): Pauli errors flip an even number of black links, on which the
\(\varepsilon\) live. They can be written as products of Wilson lines which
end on the \(\varepsilon\) (see \cref{fig:tHooft-line-boundary} and
\cite{LENSKY2023169286}).

Isolated \(\varepsilon\) (away from boundaries and \(\sigma\)) are detected
as pairs of \(\ZIIS\) fluxes. Single fluxes can be separated by utilizing a
standard object in lattice gauge theories: the 't Hooft line. They are lines
through the dual lattice, and formally correspond to flips of the gauge
fields on the links they pass through. The fermion parity constraint forces
't Hooft lines to pass through an even number of links to be gauge-invariant.
They are operators which create single-flux excitations at their ends, which
are in some contexts called \(e,m\) excitations.
More generally, a state with only a pair of separated \(\ZIIS\) fluxes is just
a state with many fermions along the 't Hooft line connecting them.

As we discussed in \cref{sec:majoranas-as-macro-dof}, we optimize for logical
fidelity by first optimizing the code distance. We now discuss the
computation of code distance in terms of Wilson and 't Hooft lines. Formally,
code distance is the minimal weight of the logical operator. The motivation
of such a definition is that such an operator produces an undetectable
logical error. The important logical operators in our case have two
interpretations: \(\varepsilon\) transfer between two \(\sigma\), typically
by Wilson lines, and 't Hooft loops around even numbers of \(\sigma\).

Wilson lines that end on an undimerized Majorana \(\sigma\)
just flip the fusion channels involving \(\sigma\); microscopically, this is
because there is no black edge touching \(\sigma\). Wilson lines between the
Majoranas generate the logical Paulis, and therefore the shortest Wilson
lines between the undimerized Majorana fermions furnish an upper bound on the
support of the lowest weight logical operator. We think of such undetectable
errors as a transfer of \(\varepsilon\) between Majorana fermions.

The total fusion outcome parity of an even number of Majoranas can also be
detected by braiding a \(\mathbb{Z}_2^{(S)}\) flux around the group. The
shortest such braids are usually generated by 't Hooft loops. These are
typically longer than the shortest Wilson lines in the bulk. In practice, 't
Hooft lines are most important in the presence of boundaries, where one also
needs to check the 't Hooft lines which start and end on boundaries and
separate the Majorana fermions into even nonempty subsets.

\begin{figure}[t]
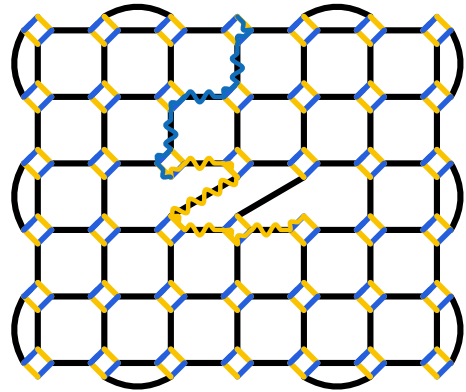

  \centering
  \include{figures/kk-3qubit-memory-wilson-lines}
  \caption{Majorana graph for a small instance of dense memory. Blue and yellow lines show Wilson line undetectable errors, see Fig.~\ref{fig:small-memories}(c).}
  \label{fig:3q-dense-mj-graph}
\end{figure}

For many common graph geometries, it is sufficient to check only a few
special cases to bound the code distance. For bulk Majoranas (i.e. away from
any boundary) the lowest weight Pauli strings flipping its fusion channels
are indeed Wilson lines, even if they are connecting to Majoranas near a
boundary. For Majoranas near a boundary, one should also check 't Hooft
lines, see example Fig.~\ref{fig:tHooft-line-boundary}. This is the sense in which the lowest weight of an undetectable error
is set by Majorana separation. An immediate consequence of these rules is the
pair of metrics for Wilson and 't Hooft lines used for the square grid in
\cref{sec:majoranas-as-macro-dof}. A detailed example for the dense memory
\cref{fig:dense-3q-ir-mjs} is shown in \cref{fig:3q-dense-mj-graph}.

To summarize, we are able to understand the properties of a timeslice of the
Majorana codes solely by the positions of the undimerized Majoranas. The role
of microscopics is only in the details of how to measure the lengths of
Wilson lines and single-flux trajectories. It is straightforward to draw some
representative Majorana graphs for given positions of the Majorana fermions
and boundaries. For any particular graph it is efficient to compute the
shortest Wilson lines and logical 't Hooft lines. To first approximation, for
locally square graphs we take a Manhattan distance aligned with the graph to
set the lengths of Wilson lines. For 't Hooft lines, we also take a Manhattan
distance in the same setting, but now between the appropriate plaquettes.

\section{Moving Majorana fermions fault-tolerantly}
\label{sec:moving-major-ferm}

Turning \cref{fig:braid-outline} into a fault-tolerant gate requires
operations to move the Majoranas, described in more detail below, as well as
circuits to measure the stabilizers as the Majoranas move. It is therefore
important to understand how to choose concrete circuits for moving Majorana
fermions and measuring the stabilizers, while maintaining code distance.

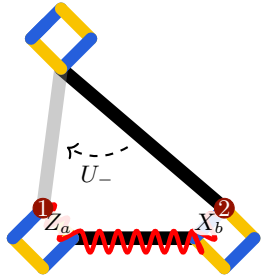
\begin{figure}[htbp!]
  \centering
  \input{figures/wl-move}
  \caption{Representative Wilson line generating the motion of the Majorana
    fermion from position 1 to position 2. Adapted from
    \cite{LENSKY2023169286}. As detailed in \cite{LENSKY2023169286}, this
    Wilson line should be exponentiated with particular sign to swap the
    Majorana fermions without creating \(\mathbb{Z}_2^{(S)}\) flux. The gauge invariant swap corresponding to \cref{eq:swap-non-gauge-invariant-unitary} in this case is \(\exp \left( i \frac{\pi}{4} Z_a X_b \right)\).}
  \label{fig:isentropic-swap}
\end{figure}

First, we point out that there are two natural ways to move anyons. One is
simply by a unitary which swaps particle positons; for lattice Majorana
fermions, for the swap \(\alpha \leftrightarrow \beta\)
\begin{equation}
  \label{eq:swap-non-gauge-invariant-unitary}
  U_{\pm} = \exp \left( \mp \frac{\pi}{4} \alpha \beta \right),
  \; U_{\pm} \beta U_{\pm}^{\dagger} = \mp \alpha,
  \; U_{\pm} \alpha U_{\pm}^{\dagger} = \pm \beta.
\end{equation}
For our Majorana fermions, gauge invariant swaps are actually generated by
the Wilson line connecting these Majorana fermions, as shown in
\cref{fig:isentropic-swap}. They are a transition from one Majorana graph to
another, and will not create \(\mathbb{Z}_2^{(S)}\) fluxes in the absence of
errors\footnote{In the absence of coherent errors it is not important that we
  make the choice that does not create flux, only that we know whether or not
  we did.}.
This sort of anyon motion (and indeed unprotected braiding) has been
demonstrated experimentally \cite{lenskygvbraiding2023}. We refer to this as ``isentropic''
motion. All that needs to be done to make this process protected is to
construct circuits to measure stabilizers concurrently with the motion. This
can be done immediately after each Majorana fermion is moved to the desired
location, or merged with the motion circuit itself.

\begin{figure*}[t]
  \centering
  \newcommand{\savedpicture}[2]{%
    \newsavebox{#1}
    \begin{lrbox}{#1}\input{#2}\end{lrbox}
  }
  \newsavebox{\VMFinalState}
  \begin{lrbox}{\VMFinalState}\input{figures/virtual-particle-final-state}\end{lrbox}

  \newsavebox{\VMWilsonLines}
  \begin{lrbox}{\VMWilsonLines}\input{figures/virtual-particle-move}\end{lrbox}

  \begin{tikzpicture}
    \node (wilson-lines) at (0, 0) {\usebox{\VMWilsonLines}};
    \node (particles-left) at (22.5em, 0) {\usebox{\VMFinalState}};
    \draw[->, line width=.1em] (wilson-lines) -- (particles-left);
  \end{tikzpicture}
  \caption{Virtual particle motion for Majorana fermions. We move the
    Majorana fermion at the top middle to the location indicated by the
    dashed line. The blue lines indicate the Majorana fermion pairs created
    from vacuum, \(1\). They are fused in the pairs indicated by the red
    lines. These fusion outcomes are random; on the right, we show a scenario
    where two of the fusions result in \(\varepsilon\).}
  \label{fig:mj-virtual-motion}
\end{figure*}
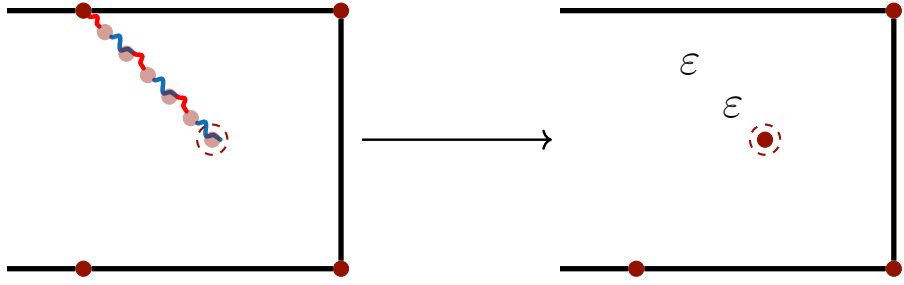

There is another way to move anyons, familiar from the Abelian case and
mentioned above. As shown in \cref{fig:mj-virtual-motion}, one way to move an
anyon is to pair-create many particle-antiparticle pairs along its path. This
can be done in \(O(1)\) time. Then, one can annihilate the pairs, offset by 1
site.  If all the pairs fuse to the vacuum, we have successfully moved the
particle to the endpoint of the path. This is always the case for Abelian
anyons. For Majoranas, we may find \(\varepsilon\) or \(1\), with equal
probability, left as outcomes of the fusion along the path. Rather than
giving up if we find \(\varepsilon\), we can simply fuse them back into the
Majorana at the end of the chain. Then the process is the same as though all
the fusion outcomes had been the vacuum. This implements the motion from the
start to the end of the chain. We comment on a few practical details. One may
also wonder how it is that we can move a particle that carries information
apparently instantaneously. There are two caveats. One is that we must
communicate the fusion outcomes over at least the same distance (this
highlights the difference to Abelian anyons, which can move instantly but
carry no information). Additionally, in the presence of measurement errors we
must make repeated measurements of the fusion outcomes (scaling with \(d\))
to maintain code distance and performance.

Using the length calculations of the previous section we see that virtual
particle motion has the same distance properties as isentropic
motion. Namely, one can just follow the paths (e.x. \cref{fig:braid-outline})
of isentropic motion, and if those paths are distance preserving, so will be
a virtual particle motion that follows the same paths. For this conclusion to
be valid, it is crucial that during the virtual particle motion it is
specifically pairs of \(\ZIIS\) fluxes, \(\varepsilon\), that are randomly
created.

\section{Syndrome extraction for Majorana graphs}
\label{sec:ziis-flux-meas}

\begin{figure}[htbp!]
  
  \noindent
  \subfloat[4 qubit stabilizer measurement.\label{fig:std-4q-measurement}]{
        \input{figures/mj-4q-stab}
  } \hfill
  \subfloat[5 qubit stabilizer measurement.\label{fig:5q-restoring-measurement}]{
    \input{figures/mj-5q-stab}
  } \hfill
  \subfloat[6 qubits stabilizer measurement.\label{fig:6q-restoring-measurement}]{
    \input{figures/mj-6q-stab}
  }
  \hfill
  \caption{Example stabilizer measurement circuits for Majorana graphs on
    devices with square grid connectivity and only certain ``ancilla'' qubits
    that can be measured (and when one ancilla is measured, all ancillae must
    be measured). In \cref{fig:std-4q-measurement}, to introduce the notation
    we show the standard circuit for a surface code plaquette. Each gate is
    indicated by a line with two bars. The number in the center says in which
    parallel layer of gates that particular gate occurs. The colors of the
    bars (see \cref{fig:mj-gauge-embedding}) indicate which controlled Pauli
    gate is done. Explicitly, if a gate goes between qubit \(a,b\) with Pauli
    colors corresponding to \(\sigma_a, \sigma_b\), the gate corresponding to
    the bar is
    \(\exp \left[ i \frac{\pi}{4} (1 - \sigma_a) (1 - \sigma_b)
    \right]\). For example, all the gates in \cref{fig:std-4q-measurement}
    are \(\mathsf{CNOT}\). The circuit is interpreted as using the ancilla
    qubit as a register for an addition operation from each qubit in the
    stabilizer. In \cref{fig:5q-restoring-measurement}, we show a 5-cycle
    measurement of a 5-qubit stabilizer. The circuit is constructed by using
    the right ancilla as a register for the ``value'' of the upper right
    qubit, transfering it to the central qubit, and finally to the left
    ancilla. Then the additions into the right ancilla are undone by applying
    the gates in reverse order. This circuit commutes with the original
    5-qubit stabilizer, which can be desireable for technical convenience. If
    we do not require the Majorana graph to return precisely to its original
    structure (but still measure all ancillas), the measurement can be done
    in 4 layers by excluding the gate indicated by the red rectangle. In
    either case, measuring the left ancilla gives the value of the
    stabilizer, while measuring the right ancilla flags an error. We
    emphasize that these circuits and gate orderings are just specific
    examples; to optimize code distance, depth, and eventually logical error
    rate the choice of circuit and gate ordering must be made carefully. We
    illustrate the flexibility of such constructions to make these
    optimizations by arranging that the gates are applied between the left 4
    stabilizer qubits and left ancilla in the same relative order.}
  \label{fig:general-stab-mu-circuits}
\end{figure}
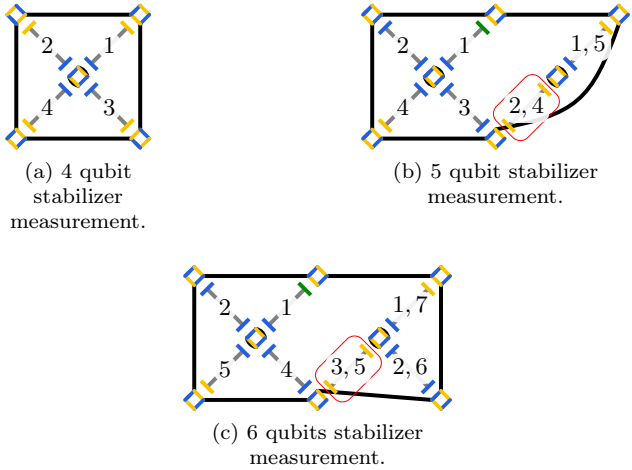

Next, we comment on the problem of stabilizer measurement: we must measure
stabilizers to maintain fault-tolerance (and additional measurements to learn
the fusion outcomes if we are using virtual particle motion), both in static
configurations and as Majorana fermions are moving. Circuits to measure
square stabilizers with square grid connectivity are commonly used in the
surface code. As can be seen from \cref{fig:3q-dense-mj-graph}, undimerized
bulk Majorana fermions result in non-square Majorana graphs with stabilizers
of weight greater than four. Finite depth circuits for measuring stabilizers
of a general shape for a device with square grid connectivity are shown in
\cref{fig:general-stab-mu-circuits}. We aim to evaluate the effect of syndrome
extraction circuits for such stabilizers on the code distance.

Let us assume we have computed the distances of the paths and static
configurations, as described in
\cref{sec:majoranas-as-macro-dof,sec:majorana-fermions-as-micro-dof}, to be
\(d\).  Given a local error model and connectivity, the existence of
\emph{some} finite-depth combined motion and syndrome extraction circuit for
the braids above, such that the \emph{process} has \emph{some} finite
distance \(\propto d\), follows from the general considerations of
\cref{sec:majoranas-as-macro-dof,sec:majorana-fermions-as-micro-dof,sec:moving-major-ferm}. It
is essentially a consequence of the locality (and therefore finite weight) of
stabilizers.

\begin{figure}[t]
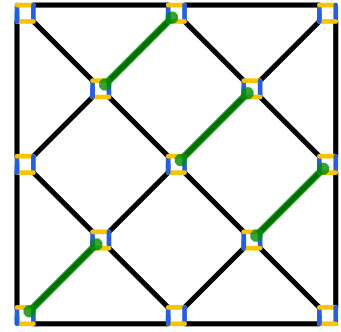

  \centering
  \include{figures/fermion-error-sc3}
  \caption{Fermion error analysis. The support of an error channel immediately after a 2-qubit gate layer is shown in green.}
  \label{fig:fermion-error-analysis}
\end{figure}

To prove that the process distance is actually \(d - O(1)\) (rather than just
\(\propto d\)) requires accounting for many details of a particular device
(most importantly the error model and connectivity), as well as error
propagation under specific stabilizer measurement and motion circuits. There
are methods for optimizing such circuits to particular hardware constraints
and error models. One can also analytically prove existence of circuits that
do not reduce code distance under weak assumptions. We leave the discussion
of these general methods to a future work. Here, we illustrate the utility of
the Majorana-based approach in a specific example. Many error models have
2-qubit noise channels applied to qubits after they undergo 2-qubit gates. As
shown in \cref{fig:fermion-error-analysis}, if each 2-qubit error counts as a
single elementary error such an error pattern can cut the effective length
of a Wilson line, and therefore the code distance, in half. The pattern shown
in \cref{fig:fermion-error-analysis} appears in even the standard surface
code syndrome extraction circuit; it is not distance reducing since the
Majorana fermions at opposite corners are twice as far from each other as
they are to their nearest neighbors. Circuits with Majorana fermions in the bulk must
be designed to avoid such errors to avoid cutting the code distance by some
fraction. The mitigation from the perspective of the Majorana graph is clear;
one must design circuits such that intermediate Majorana graphs do not have Wilson
lines aligned with 2-qubit gate layers as in
\cref{fig:fermion-error-analysis}. We employ this mitigation in our circuits
and do not suffer fractional distance reductions due to such errors.

In \cref{sec:numerics}, we show an existence proof for a system with several
realistic constraints: concrete circuits for an architecture inspired by
superconducting processors with square-grid connectivity and where only
certain ``ancilla'' qubits have readout and reset operations. These circuits
have a code distance \(d\) when implemented using \(4d^2 + O(d)\) qubits.

\section{Asymptotic comparison to lattice surgery}
\label{sec:gates-braiding}

A practical figure of merit for a gate protocol is the logical gate fidelity
at fixed physical resources. In particular, we fix the number of physical
qubits and device error model\footnote{As discussed below, we actually add
  additional restrictions, such as readout only on certain ``ancilla'' qubits
  and square grid connectivity, which are already satisfied by lattice
  surgery.} So far, we have given results in terms of the code distance.
The code distance is a useful proxy since fidelities are expected to have the
leading (in code distance, or equivalently number of qubits) behaviour
\begin{equation}
  \label{eq:asymptotic-ec-scaling}
  \mathsf{LER} \sim c[d(n)] \Lambda^{-\lfloor \frac{d(n)+1}{2} \rfloor}
\end{equation}
for some model-dependent constant \(\Lambda\) and subleading behaviour
captured by \(c\). In this work, the logical error rate \(\mathsf{LER}\)
refers to \(1 - \mathcal{F}\), where \(\mathcal{F}\) is the average-state
gate fidelity.
The parameter \(\Lambda\) and function \(c\) are sensitive to the details of
the gate schedules. To see if there is a scaling advantage for a particular
family of devices and error models between two protocols, we would have to
compute (or bound) \(\Lambda\) for each protocol. We leave this to future
work.

\begin{figure*}[t]
  \centering

  \subfloat{\input{figures/lattice-surgery}} \qquad \qquad
  \subfloat{\includegraphics{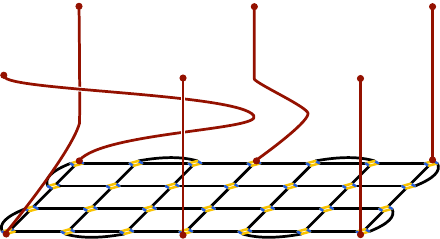}}
  \caption{Left: qubit layout for lattice surgery. Right: qubit layout and Majorana fermion trajectories for braiding.}
  \label{fig:ls-vs-braiding}
\end{figure*}

Here, we simply point out that if one is comparing two protocols \(\alpha, \beta\) with distances \(d_{\alpha}(n) > d_{\beta}(n)\), a natural figure of merit is
\begin{equation}
  \eta_{\beta \to \alpha} = \lim_{n \to \infty} 1 - d_{\beta}(n) / d_{\alpha}(n).
\end{equation}
The first-order expectation is that since \(\eta_{\beta \to \alpha} > 0\), it
is advantageous to use protocol \(\alpha\). One may ask how the details of
protocol implementation affect relative performance. The local differences
between protocols (number of layers between measurement cycles, stabilizer
size, etc) are captured by the ratio \(\Lambda_{\alpha} /
\Lambda_{\beta}\). For many error models, this ratio is at most weakly
dependent on the scales of local error rates. We have that
\begin{gather}
 \mathsf{LER}_{\alpha} / \mathsf{LER}_{\beta} \sim \left[ (\Lambda_{\alpha}
  / \Lambda_{\beta}) \Lambda_{\beta}^{\eta_{\beta \to \alpha}}
\right]^{-d_{\alpha}(n) / 2},
\end{gather}
so there is a scaling advantage to protocol
\(\alpha\) (which grows exponentially in code distance) as long as
\begin{gather}
\Lambda_{\alpha} / \Lambda_{\beta} > \Lambda_{\beta}^{- \eta_{\beta \to
    \alpha}}.
\end{gather}
Note this condition itself becomes exponentially weaker as
the device parameters improve, so one can afford to lose quite a significant
amount of \(\Lambda_{\alpha}\).

A leading protocol for gates in surface-code encoded qubits is lattice
surgery. In these scheme, the only fundamental operations at the logical
level are initialization and Pauli measurement. This uses only the fusion
algebra of the Majorana fermions. Such a scheme requires a logical ancilla to
perform unitary gates. The corresponding description for the protocol shown
here is that the fundamental logical operations are all unitary, which we
achieve by using the braiding algebra of the Majorana fermions. The space
advantage (at fixed distance) of the braiding protocol shown here arises
simply from the detailed analysis of the Majorana fermions presented above,
and rests on two key aspects. We remove the need for a logical ancilla by
using the braiding statistics, and pack the braid as tightly as possible
according to the Majorana distance discussed in
\cref{sec:majoranas-as-macro-dof}. The final result is that a distance \(d\)
gate is achievable with \(4d^2 + O(d)\) qubits by braiding, while lattice
surgery requires \(6d^2 + O(d)\). This means
\begin{equation}
  \eta_{\mathsf{LS} \to \mathsf{B}} = 1 - \sqrt{2/3} \approx 0.184.
\end{equation}
We will show numerical evidence that for certain error models and schedules
\begin{equation}
  \label{eq:ls-comparison-equation}
  \Lambda_{\mathsf{B}} / \Lambda_{\mathsf{LS}} > \Lambda_{\mathsf{LS}}^{- \eta_{\mathsf{LS} \to \mathsf{B}}},
\end{equation}
and moreover that the functions \(c\) are such that even naive
implementations of a braiding protocol are advantageous already at small code
distances.

\section{Numerics}
\label{sec:numerics}

We move on to numerical benchmarks of a simple implementation of a
fault-tolerant braiding gate: we follow steps 2-4 of the path
\cref{fig:braid-outline}. This is a typical gate, since the Majorana
configuration is the same at the start and end. If this gate were part of a
larger circuit, we may also need to perform one or both of steps 1 and 5, but
for the minimal example we do not. We utilize virtual particle motion for
technical convenience. In order to showcase both the flexibility of our
approach and the suitability for near-term devices, we consider realistic,
but rigid constraints on connectivity and qubit operations. Our device model
is motivated by the recent realization of beyond break-even surface code
memory in a similar setting \cite{acharyaQuantumErrorCorrection2025}. In
particular, we take nearest-neighbor connectivity on the square lattice for
2-qubit gates. We only allow measure and reset operations on the \(A\)
sublattice, and call these the ``ancilla'' qubits. Moreover, on every
measurement \emph{all} ancilla qubits must be measured.

  \begin{figure*}
    \centering
    \includegraphics[width=0.7\textwidth]{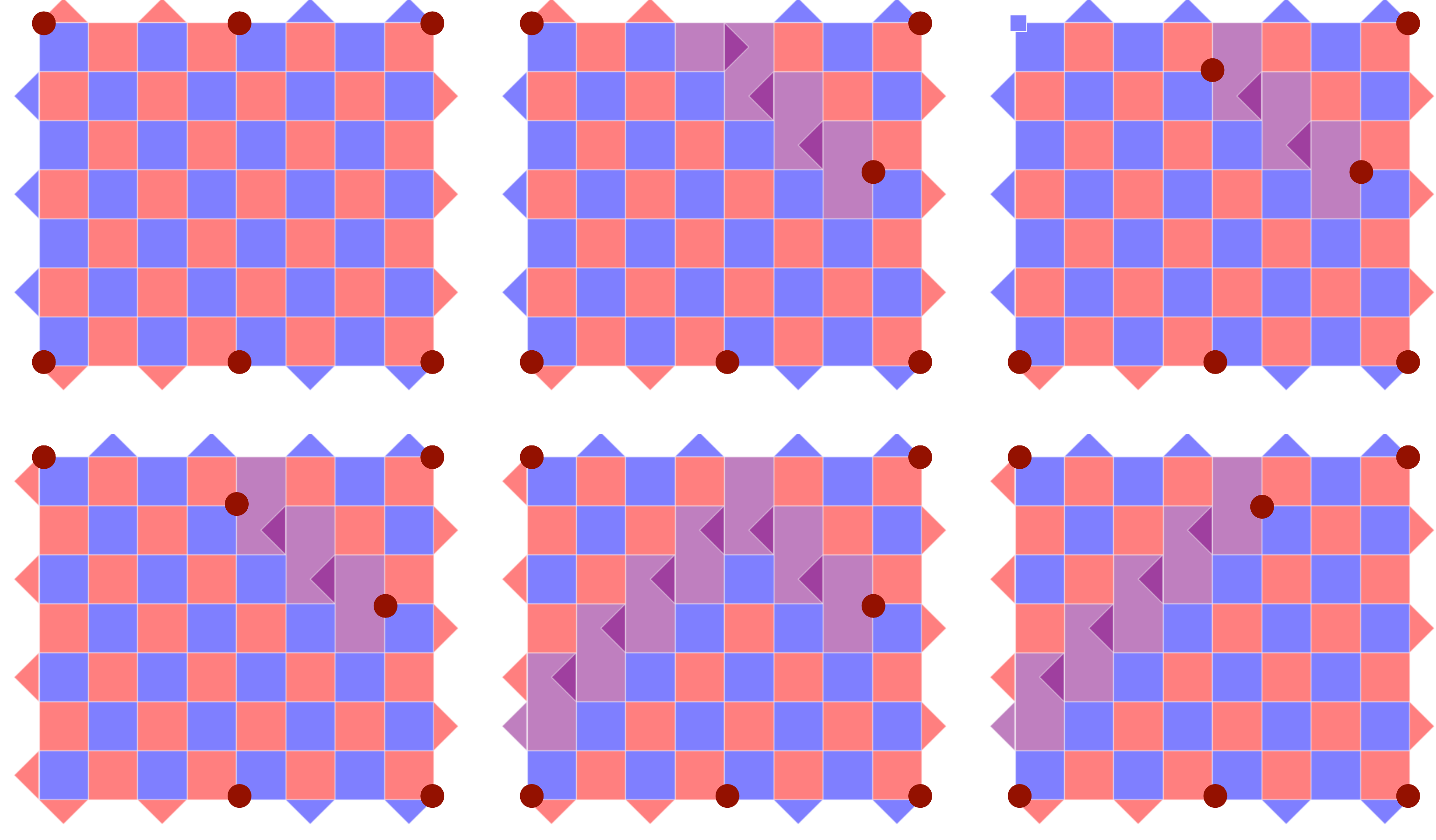}
    \caption{Stabilizer configurations of an unoptimized virtual
      particle-based braid, following steps 1-5 of
      \cref{fig:braid-outline}. Time order is left-to-right, top to
      bottom. Majorana locations are indicated with red dots. Triangular
      stabilizers are 2-qubit stabilizers at the vertices of their long
      end. A simple circuit to implement a braid using this sequence of codes is benchmarked in \cref{fig:braiding-ls-numerics}.}
    \label{fig:unoptimized-stab-configs}
  \end{figure*}

We consider an error model called \SITK{}. The details of this error model
are in \cite{Gidney2021faulttolerant}. At a high level, after every 2 qubit
operation we apply a depolarizing channel on those 2 qubits with some
parameter. After every 1 qubit operation we apply a depolarizing channel on
that qubit. The circuit is divided into parallel layers, and in each layer
the identity operation counts for a single-qubit operation (this is called
``idle'' error).

To provide a baseline, we compare to a minimal gate by lattice surgery under
the same error model, with the same device constraints. Note that we chose
our constraints so that they are already satisfied by standard lattice
surgery circuits. The comparison is done by sampling errors using STIM
\cite{gidney2021stim} and decoding by PyMatching
\cite{Higgott2025sparseblossom}.

\begin{figure*}
  \centering
  \includegraphics[width=\textwidth]{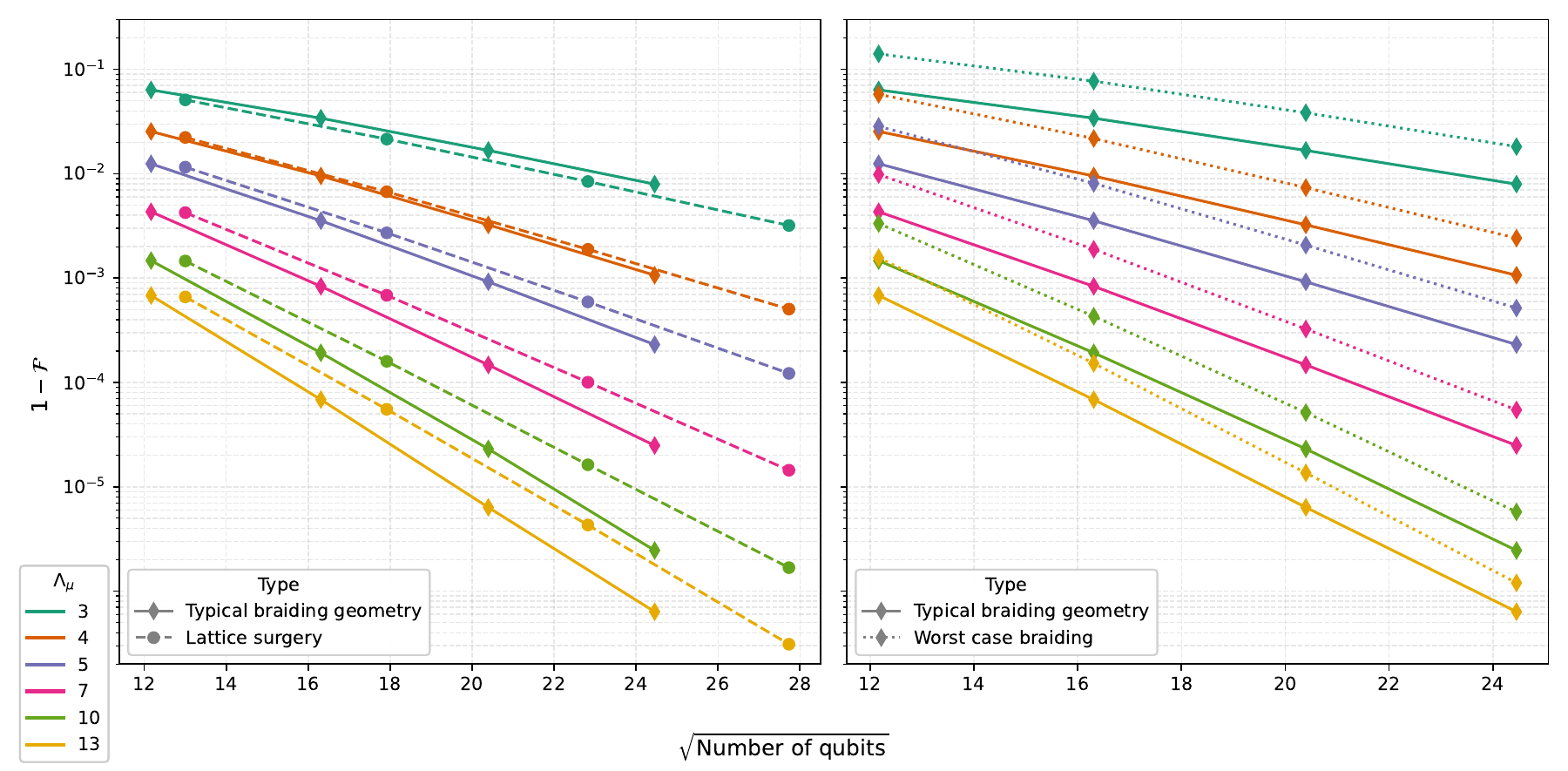}
  \caption{The gate fidelity (or \(\mathsf{LER}\)) for a simple virtual
    particle-based braiding gate (square points) as a function of the (square
    root) of the number of qubits. Different colors correspond to different
    error rate parameters in \SITK{}. We summarize the rate parameters by
    numerically computing \(\Lambda_{\mu}\) from
    \cref{eq:asymptotic-ec-scaling} for a standard surface code memory
    experiment using the same error model and decoder. We recover the common
    relationship that \(\Lambda_{\mu} = p_0 / p\), where \(p\) is the
    parameter of \SITK{} and \(p_0 \approx 5.3 \times 10^{-3}\). Lines are a
    guide to the eye. Lattice surgery error rates are shown
    using circles.}
  \label{fig:braiding-ls-numerics}
\end{figure*}

Stabilizer configurations after each virtual particle motion step
\cref{fig:mj-virtual-motion} are shown in
\cref{fig:unoptimized-stab-configs}. These are not optimal stabilizer
configurations, but lend themselves to simple stabilizer measurement
circuits. The logical error rates shown in \cref{fig:braiding-ls-numerics}
provide numerical confirmation for the theoretical discussions of the
previous sections, as well as estimates of comparison to lattice surgery for
small devices. Although we have not optimized the circuits or stabilizer
configurations, we already find evidence \cref{eq:ls-comparison-equation} is
satisfied, so that there is an exponential scaling advantage in logical
fidelity. Moreover, we see that the performance even at small sizes favors
braiding as well for certain operations and low enough error rates.

One may wonder how would the most resource demanding gate by braiding compare
to the typical. In the geometry of Fig.~\ref{fig:braid-outline} the swap of
two Majorana fermions at the boundary is such most demanding gate corresponds
to steps 1-5. Comparison of the two gates is shown in
Fig.~\ref{fig:braiding-ls-numerics}.

\section{Conclusion}
\label{sec:conclusion}

In conclusion, we demonstrated that macroscopic Majorana degrees of freedom
are a powerful and versatile tool for the analysis of the local and global
building blocks of fault-tolerant quantum computation in 2D. It allows for
optimization and code distance analysis of fault-tolerant memories and gates.

We studied and microscopically justified macroscopic properties of Majorana
graphs. Using our approach, we designed a fault-tolerant protocol for the
logical 2-qubit Clifford group. Macroscopically, the fundamental primitives
and sources of advantage for our protocols are the braiding statistics and
packing properties of Majorana fermions. Lucid connection of our method
to geometry and packing of Majorana fermions makes the interdependence between
dense memory and braiding immediate. Leveraging the dense packing gives a
dense memory; more importantly, we use this example to illustrate that there
is room to braid the Majorana fermions in the space of 2 surface code
patches. To understand the performance of such gates, we argue that they
outperform lattice surgery for a wide range of error models and other device
constraints. Numerically, we confirm that even unoptimized protocols on
constrained devices are superior at large system sizes, and find advantage
even for small systems.

Our construction provides a basis for analysis of Pauli codes, both in 2 and
higher dimensions (for example reconfigurable atom arrays or modular
superconducting architectures). More generally, we expect application of our
topological approach to dynamic and static codes will open a new route for
the co-design of fault-tolerant protocols, quantum computer architectures,
and compilation of fault-tolerant computations (say to braids and fusions).

\bibliography{./includes/old-nab-refs.bib} 

\appendix

\include{SupplementaryMaterial.tex}

\end{document}

%% file: figures/graphical-notation-sc-mjs.tex
\begin{tikzpicture}
  \node[rectangle, inner xsep = 1cm, inner ysep = 1cm] (rct) at (0, 0) {};
  \begin{scope}[every node/.style={mj indicator, minimum size=.2cm}]
    \foreach \nn/\npos in {nw/north west, sw/south west, ne/north east, se/south east} {
      \node (\nn) at (rct.\npos) {};
    }
  \end{scope}
  \begin{scope}[every path/.style={line width=.07cm}]
    \draw (nw) -- (ne);
    \draw (sw) -- (se);

    \draw (nw) -- (sw);
    \draw (ne) -- (se);
  \end{scope}
  \begin{scope}[every path/.style={wilson line, line width=.2em}]
    \draw[mjx] (sw) -- (nw);
    \draw[mjy] (sw) -- (ne);
    \draw[mjz] (sw) -- (se);
  \end{scope}
  \begin{scope}[every node/.style={
      fill=white, fill opacity=0.9, text opacity=1, inner sep=.1em,
    rounded corners=.1em}]
    \node at ($(sw)!.5!(nw)$) {\(X\)};
    \node at ($(sw)!.5!(se)$) {\(Z\)};
    \node at ($(sw)!.5!(ne)$) {\(Y\)};
  \end{scope}
\end{tikzpicture}

%% file: figures/6-mj-mjs.tex
\begin{tikzpicture}
  \node[rectangle, inner xsep = 2cm, inner ysep = 1cm] (rct) at (0, 0) {};
  \begin{scope}[every node/.style={mj indicator, minimum size=.2cm}]
    \foreach \nn/\npos in {nw/north west, sw/south west, ne/north east, se/south east, nc/north, sc/south} {
      \node (\nn) at (rct.\npos) {};
    }
  \end{scope}
  \begin{scope}[every path/.style={line width=.07cm}]
    \draw (nw) -- (nc);
    \draw (sw) -- (sc);

    \draw (nw) -- (sw);
    \draw (nc) -- (ne);
    \draw (sc) -- (se);
    \draw (ne) -- (se);
  \end{scope}

  \begin{scope}[every path/.style={-Stealth, thick}]
    \draw (sw) to[bend left] (nw);
    \draw (nw) to[bend left] (sw);
  \end{scope}

  \begin{scope}[every path/.style={wilson line, line width=.2em}]
    \draw[mjx] (nw) -- (nc);
    \draw[mjx] (sw) -- (sc);
  \end{scope}

  \begin{scope}[every node/.style={
      fill=white, fill opacity=0.9, text opacity=1, inner sep=.1em,
      rounded corners=.1em}]
    \node at ($(nw)!.5!(nc)$) {\(X_1\)};
    \node at ($(sw)!.5!(sc)$) {\(X_2\)};
  \end{scope}
\end{tikzpicture}

%% file: figures/3q-mj-mjs.tex
\begin{tikzpicture}
  \node[rectangle, inner xsep = 2cm, inner ysep = 1cm] (rct) at (0, 0) {};
  \begin{scope}[every node/.style={mj indicator, minimum size=.2cm}]
    \foreach \nn/\npos in {nw/north west, sw/south west, ne/north east, se/south east, nc/north, sc/south} {
      \node (\nn) at (rct.\npos) {};
    }
    \coordinate (marker-left) at ($(rct.north west)!.5!(rct.north)$);
    \coordinate (marker-right) at ($(rct.north)!.5!(rct.north east)$);
    \node (cw) at (rct.west -| marker-left) {};
    \node (ce) at (rct.west -| marker-right) {};
  \end{scope}
  \begin{scope}[every path/.style={line width=.07cm}]
    \draw (nw) -- (nc);
    \draw (sw) -- (sc);

    \draw (nw) -- (sw);
    \draw (nc) -- (ne);
    \draw (sc) -- (se);
    \draw (ne) -- (se);
  \end{scope}
  \draw[wilson line, mjx] (nc) -- (cw);
  \draw[wilson line, mjz] (cw) -- (ce);
\end{tikzpicture}

%% file: figures/braiding-mj-picture.tex
\begin{tikzpicture}
  \node[rectangle, inner xsep = 2cm, inner ysep = 1cm] (rct) at (0, 0) {};
  \begin{scope}[every node/.style={mj indicator, minimum size=.2cm}]
    \foreach \nn/\npos in {nw/north west, sw/south west, ne/north east, se/south east, nc/north, sc/south} {
      \node (\nn) at (rct.\npos) {};
    }
    \coordinate (marker-left) at ($(rct.north west)!.5!(rct.north)$);
    \coordinate (marker-right) at ($(rct.north)!.5!(rct.north east)$);
    \coordinate (cw) at (rct.west -| marker-left) {};

    \node[stroke=dotted,fill=none,dashed,thick] (ce) at (rct.west -| marker-right) {};
  \end{scope}
  \begin{scope}[every path/.style={line width=.07cm}]
    \draw (nw) -- (nc);
    \draw (sw) -- (sc);

    \draw (nw) -- (sw);
    \draw (nc) -- (ne);
    \draw (sc) -- (se);
    \draw (ne) -- (se);
  \end{scope}
  \draw[thick,-{Stealth}] (nc) to [bend left] node[pos=0.7, above right]{1} (ce);
  \draw[thick,-{Stealth}] (nw) to [bend left] node[midway, above]{2} (nc);
  \draw[thick,-{Stealth}] (sw) to [bend left] node[midway, left]{3} (nw);

  \draw[thick,-{Stealth}] (nc) -- node[midway,below right] {4} (sw);
  \draw[thick,-{Stealth}] (ce) to [bend left] node[midway, below left] {5} (nc);
\end{tikzpicture}

%% file: figures/gauge-theory-outline.tex
  \begin{tikzpicture}[scale=0.8,
    transition-arrow/.style={->, line width=1.3pt},
    physical-qubit/.append style={minimum size=.4cm},
    majorana diamond edge/.append style={line width=2.5},
    plabel/.style={black, fill=white, fill opacity=.7, inner sep=2, rounded corners, text opacity=1}]
    \def\tetralen{1.7}
    \pgfmathsetmacro{\tetrang}{180-asin(2 * sqrt(2)/3)}
    \pgfmathsetmacro{\tetrarad}{sqrt(3/2)/2}
    \pgfmathsetmacro{\tetrarot}{45}
    \coordinate (qubit) at (0,0,0) {};
    \begin{scope}
      \begin{scope}[every node/.style={majorana vertex}, scale=\tetralen, shift=(qubit)]
        \foreach \pn in {1,2,3} {
          \node[text=white] (mj-\pn) at ($({sin(\tetrang) * cos((\pn-1)*120 + \tetrarot)}, {cos(\tetrang)}, {sin(\tetrang) * sin((\pn-1)*120 + \tetrarot)})$) {};
        }
        \node[text=white] (mj-4) at (0, 1, 0) {};
      \end{scope}
      \draw[majorana diamond edge, rev kasteleyn arrow, tau3link, dashed] (mj-2) -- (mj-3);
      \node[physical-qubit, minimum size=.4cm] (qubit) at (qubit) {};
      \begin{scope}[every path/.style={line width=1pt, ->, shorten >= 1pt, shorten <= 1pt}]
        \foreach \x in {1,...,4} {
          \draw (qubit) -- (mj-\x);
        }
      \end{scope}
      \begin{scope}[every path/.style={majorana diamond edge}, local bounding box=tetrabox]
        \draw[rev kasteleyn arrow, mjz] (mj-1) -- node[midway, plabel, below] {\(Z\)} (mj-2);
        \draw[kasteleyn arrow, mjz] (mj-3) -- node[midway, plabel, right] {\(Z\)} (mj-4);
        \draw[kasteleyn arrow, mjx] (mj-1) -- node[midway, plabel, right] {\(X\)} (mj-3);
        \draw[kasteleyn arrow, mjx] (mj-2) -- node[midway, plabel, left] {\(X\)} (mj-4);
        \draw[kasteleyn arrow, mjy] (mj-1) -- (mj-4);
      \end{scope}
    \end{scope}
    \node[physical-qubit] (initial-qubit) at ($(tetrabox.west) + (-2,0)$) {};
    \node[below=.3] at (initial-qubit) {qubit};
    \draw[transition-arrow, shorten <= 1] (initial-qubit) -- (tetrabox.west);
    \pic[mj diamond size=1.3cm] (diamond) at ($(tetrabox.east)+(2.5,0)$) {flat pf diamond};
    \begin{scope}[every node/.style={plabel, fill opacity=0.8}]
      \node at ($(diamond-dmnd.east)!.5!(diamond-dmnd.north)$) {\(Z\)};
      \node at ($(diamond-dmnd.west)!.5!(diamond-dmnd.north)$) {\(X\)};
      \node at ($(diamond-dmnd.west)!.5!(diamond-dmnd.south)$) {\(Z\)};
      \node at ($(diamond-dmnd.south)!.5!(diamond-dmnd.east)$) {\(X\)};
    \end{scope}
    \draw[transition-arrow] ($(tetrabox.east)+(0.4, 0)$) -- ($(diamond-dmnd.west)+(-0.3,0)$);
    \begin{scope}[mj diamond size=20, local bounding box=single-edge-bb]
      \pic (d1) at ($(diamond-dmnd) + (3,1.2)$) {flat pf diamond};
      \pic (d2) at ($(diamond-dmnd) + (4,-1.2)$) {flat pf diamond};
      \draw[stabilizer edge] (d1-dmnd.south) -- (d2-dmnd.north);
      \pic (d1) at ($(diamond-dmnd) + (3,1.2)$) {flat pf diamond};
      \pic (d2) at ($(diamond-dmnd) + (4,-1.2)$) {flat pf diamond};
    \end{scope}
    \draw[transition-arrow] ($(diamond-dmnd.east) + (0.4,0)$) -- ++(.9, 0);

    \pgfdeclarelayer{stabedges}
    \pgfdeclarelayer{mjdiamonds}
    \pgfsetlayers{stabedges,mjdiamonds,main}
    \begin{scope}[xshift=27em,yshift=-7.5em, mj diamond size=10, local bounding box=surface-code-bb]
      \def\scxdist{3}
      \def\scydist{3}
      \begin{scope}[scale=1.3]
        \begin{pgfonlayer}{mjdiamonds}
          \foreach \x in {1,...,\scxdist} {
            \foreach \y in {1,...,\scydist} {
              \pic[name=qb-\x-\y] at (\x,\y) {flat pf diamond};
            }
          }
        \end{pgfonlayer}
      \end{scope}
      \begin{pgfonlayer}{stabedges}
        \begin{scope}[every path/.style=stabilizer edge]
          \foreach \x in {1,...,\scxdist} {
            \foreach \y in {1,...,\scydist} {
              \ifnum \x<\scxdist
                \pgfmathtruncatemacro{\nextx}{\x+1}
                \draw (qb-\x-\y-dmnd.east) -- (qb-\nextx-\y-dmnd.west);
              \fi
              \ifnum \y<\scydist
                \pgfmathtruncatemacro{\nexty}{\y+1}
                \draw (qb-\x-\y-dmnd.north) -- (qb-\x-\nexty-dmnd.south);
              \fi
            }
          }
          \foreach \x in {1,...,\scxdist} {
            \pgfmathtruncatemacro{\nextx}{\x+1}
            \ifnum\nextx>\scxdist\else
              \ifodd\x
                \draw (qb-\x-\scydist-dmnd.north) to[bend left]
                (qb-\nextx-\scydist-dmnd.north);
              \else
                \draw (qb-\x-1-dmnd.south) to[bend right]
                (qb-\nextx-1-dmnd.south);
              \fi
            \fi
          }
          \foreach \y in {1,...,\scydist} {
            \pgfmathtruncatemacro{\nexty}{\y+1}
            \ifnum\nexty>\scydist\else
              \ifodd\y
                \draw (qb-1-\y-dmnd.west) to[bend left]
                (qb-1-\nexty-dmnd.west);
              \else
                \draw (qb-\scxdist-\y-dmnd.east) to[bend right]
                (qb-\scxdist-\nexty-dmnd.east);
              \fi
            \fi
          }
        \end{scope}
      \end{pgfonlayer}
    \end{scope}

    \draw[transition-arrow, shorten >= 1em] (single-edge-bb.east) -- (single-edge-bb.east -| surface-code-bb.west);
  \end{tikzpicture}

%% file: figures/code-distance-tHoof-line.tex
\begin{tikzpicture}[
  mj diamond size=10,
  scale=0.8,
  plabel/.style={black, fill=white, fill opacity=.7, inner sep=2, rounded corners, text opacity=1}
]
  \pgfdeclarelayer{fluxes}
  \pgfdeclarelayer{stabedges}
  \pgfdeclarelayer{mjdiamonds}
  \pgfsetlayers{fluxes,stabedges,mjdiamonds,main}
  
  \def\scxdist{5}
  \def\scydist{5}
  
  \begin{pgfonlayer}{mjdiamonds}
    \begin{scope}[scale=1.1]
      \foreach \x in {1,...,\scxdist} {
        \foreach \y in {1,...,\scydist} {
          \pic[name=qb-\x-\y] at (\x,\y) {flat pf diamond};
        }
      }
    \end{scope}
  \end{pgfonlayer}
  
  \begin{pgfonlayer}{stabedges}
    \begin{scope}[every path/.style=stabilizer edge]
      \foreach \x in {1,...,\scxdist} {
        \foreach \y in {1,...,\scydist} {
          \ifnum \x<\scxdist
            \pgfmathtruncatemacro{\nextx}{\x+1}
            \draw (qb-\x-\y-dmnd.east) -- (qb-\nextx-\y-dmnd.west);
          \fi
          \ifnum \y<\scydist
            \pgfmathtruncatemacro{\nexty}{\y+1}
            \draw (qb-\x-\y-dmnd.north) -- (qb-\x-\nexty-dmnd.south);
          \fi
        }
      }
      
      \foreach \x in {1,4} {
        \pgfmathtruncatemacro{\nextx}{\x+1}
        \draw (qb-\x-\scydist-dmnd.north) to[bend left] (qb-\nextx-\scydist-dmnd.north);
        \draw (qb-\x-1-dmnd.south) to[bend right] (qb-\nextx-1-dmnd.south);
      }
      \foreach \y in {1,4} {
        \pgfmathtruncatemacro{\nexty}{\y+1}
        \draw (qb-1-\y-dmnd.west) to[bend left] (qb-1-\nexty-dmnd.west);
        \draw (qb-\scxdist-\y-dmnd.east) to[bend right] (qb-\scxdist-\nexty-dmnd.east);
      }
    \end{scope}
  \end{pgfonlayer}

  \begin{pgfonlayer}{main}
    \draw[mjy, line width=3pt] 
      ($($(qb-2-4-dmnd.center)!0.5!(qb-3-5-dmnd.center)$) + (0, 0.8)$) -- 
      ($(qb-2-4-dmnd.center)!0.5!(qb-3-5-dmnd.center)$) -- 
      ($(qb-3-4-dmnd.center)!0.5!(qb-4-5-dmnd.center)$) -- 
      ($(qb-3-3-dmnd.center)!0.5!(qb-4-4-dmnd.center)$) -- 
      ($(qb-4-3-dmnd.center)!0.5!(qb-5-4-dmnd.center)$) -- 
      ($(qb-4-2-dmnd.center)!0.5!(qb-5-3-dmnd.center)$) -- 
      ($($(qb-4-2-dmnd.center)!0.5!(qb-5-3-dmnd.center)$) + (0.8, 0)$);   
      
    \begin{scope}[every node/.style={plabel, fill opacity=0.8}]
      \node at ($(qb-3-5-dmnd.west)!.5!(qb-3-5-dmnd.south)$) {$Z$};
      
      \node at ($(qb-4-4-dmnd.west)!.5!(qb-4-4-dmnd.south)$) {$Z$};
      
      \node at ($(qb-5-3-dmnd.west)!.5!(qb-5-3-dmnd.south)$) {$Z$};
    \end{scope}
  \end{pgfonlayer}

  \begin{scope}[wilson line amplitude=2, wilson line segment length=3]
    \draw[wilson line, line width=0.04cm] 
    (qb-3-5-dmnd.north) 
    -- (qb-3-5-dmnd.east)
    -- (qb-4-5-dmnd.west)
    -- (qb-4-5-dmnd.south) 
    -- (qb-4-4-dmnd.north)
    -- (qb-4-4-dmnd.east)
    -- (qb-5-4-dmnd.west) 
    -- (qb-5-4-dmnd.south) 
    -- (qb-5-3-dmnd.north)
    -- (qb-5-3-dmnd.east);

    \draw[wilson line, line width=0.04,red, line width=0.04cm]
    (qb-3-1-dmnd.south) -- (qb-3-1-dmnd.north) --
    (qb-3-2-dmnd.south) -- (qb-3-2-dmnd.west) --
    (qb-2-2-dmnd.east) -- (qb-2-2-dmnd.north);
  \end{scope}

  \begin{pgfonlayer}{fluxes}
    \begin{scope}[every path/.style={blue, opacity=0.3}]
      \fill
      (qb-2-2-dmnd.north) -- (qb-2-2-dmnd.east) --
      (qb-3-2-dmnd.west) -- (qb-3-2-dmnd.north) --
      (qb-3-3-dmnd.south) -- (qb-3-3-dmnd.west) --
      (qb-2-3-dmnd.east) -- (qb-2-3-dmnd.south) --cycle;
      \fill
      (qb-1-2-dmnd.north) -- (qb-1-2-dmnd.east) --
      (qb-2-2-dmnd.west) -- (qb-2-2-dmnd.north) --
      (qb-2-3-dmnd.south) -- (qb-2-3-dmnd.west) --
      (qb-1-3-dmnd.east) -- (qb-1-3-dmnd.south) --cycle;
  \end{scope}
\end{pgfonlayer}

\node[fill=white, fill opacity=0.8, rounded corners, text opacity=1, inner sep=.24em] at ($(qb-2-2-dmnd.north)!.5!(qb-2-3-dmnd.south)$) {\(\varepsilon\)};
\end{tikzpicture}

%% file: figures/kk-3qubit-memory-wilson-lines.tex
\begin{tikzpicture}[mj diamond size=10,scale=0.8]
  \pgfdeclarelayer{stabedges}
  \pgfdeclarelayer{mjdiamonds}
  \pgfsetlayers{stabedges,mjdiamonds,main}
  \def\scxdist{7}
  \def\scydist{6}
  \begin{pgfonlayer}{mjdiamonds}
    \begin{scope}[scale=1.1]
      \foreach \x in {1,...,\scxdist} {
        \foreach \y in {1,...,\scydist} {
          \pic[name=qb-\x-\y] at (\x,\y) {flat pf diamond};
        }
      }
    \end{scope}
  \end{pgfonlayer}
  \begin{pgfonlayer}{stabedges}
    \begin{scope}[every path/.style=stabilizer edge]
      \foreach \x in {1,...,\scxdist} {
        \foreach \y in {1,...,\scydist} {
          \ifnum \x<\scxdist
            \pgfmathtruncatemacro{\nextx}{\x+1}
            \draw (qb-\x-\y-dmnd.east) -- (qb-\nextx-\y-dmnd.west);
          \fi
          \ifnum \y<\scydist
            \pgfmathtruncatemacro{\nexty}{\y+1}
            \ifnum\y=3
              \ifnum\x<3
                \draw (qb-\x-\y-dmnd.north) -- (qb-\x-\nexty-dmnd.south);
              \fi
              \ifnum\x>5
                \draw (qb-\x-\y-dmnd.north) -- (qb-\x-\nexty-dmnd.south);
              \fi
            \else
              \draw (qb-\x-\y-dmnd.north) -- (qb-\x-\nexty-dmnd.south);
            \fi
          \fi
        }
      }
      \draw (qb-3-3-dmnd.north) -- (qb-4-4-dmnd.south);
      \draw (qb-4-3-dmnd.north) -- (qb-5-4-dmnd.south);
      \foreach \x in {2,5} {
        \pgfmathtruncatemacro{\nextx}{\x+1}
        \draw (qb-\x-\scydist-dmnd.north) to[bend left]
        (qb-\nextx-\scydist-dmnd.north);
        \draw (qb-\x-1-dmnd.south) to[bend right]
        (qb-\nextx-1-dmnd.south);
      }
      \foreach \y in {1,...,\scydist} {
        \pgfmathtruncatemacro{\nexty}{\y+1}
        \ifodd\y
          \draw (qb-1-\y-dmnd.west) to[bend left] (qb-1-\nexty-dmnd.west);
          \draw (qb-\scxdist-\y-dmnd.east) to[bend right] (qb-\scxdist-\nexty-dmnd.east);
        \fi
      }
    \end{scope}
  \end{pgfonlayer}

  \begin{scope}
    \setwilsonline{%
      (qb-3-4-dmnd.south)
      -- (qb-3-4-dmnd.east)
      -- (qb-4-4-dmnd.west)
      -- (qb-4-4-dmnd.south)
      -- (qb-3-3-dmnd.north)
      -- (qb-3-3-dmnd.east)
      -- (qb-4-3-dmnd.west)
      -- (qb-4-3-dmnd.south)
      -- (qb-4-3-dmnd.east)
      -- (qb-5-3-dmnd.west)
      -- (qb-5-3-dmnd.north)%
    }
    \drawwilsonline[draw=tau2link]
  \end{scope}

  \begin{scope}
    \setwilsonline{%
      (qb-3-4-dmnd.south)
      -- (qb-3-4-dmnd.west)
      -- (qb-3-4-dmnd.north)
      -- (qb-3-5-dmnd.south)
      -- (qb-3-5-dmnd.east)
      -- (qb-4-5-dmnd.west)
      -- (qb-4-5-dmnd.north)
      -- (qb-4-6-dmnd.south)
      -- (qb-4-6-dmnd.east)
      -- (qb-4-6-dmnd.north)%
    }
    \drawwilsonline
  \end{scope}

\end{tikzpicture}

%% file: figures/wl-move.tex
\begin{tikzpicture}
  \pgfdeclarelayer{stabedges}
  \pgfdeclarelayer{mjdiamonds}
  \pgfsetlayers{stabedges,mjdiamonds,main}

  \tikzset{mj diamond size=2.4em}
  \begin{pgfonlayer}{mjdiamonds}
    \begin{scope}[scale=2.4]
      \pic (a) at (0,0) {flat pf diamond};
      \pic (b) at (1,0) {flat pf diamond};
      \pic (c) at (0.1,1.1) {flat pf diamond};
    \end{scope}
  \end{pgfonlayer}
  \begin{pgfonlayer}{stabedges}
    \begin{scope}[every path/.style={stabilizer edge}]
      \draw (a-dmnd.east) -- (b-dmnd.west);
      \draw (b-dmnd.north) -- node[midway] (start-edge-anc) {} (c-dmnd.south);
      \draw[gray, opacity=0.4] (c-dmnd.south) -- node[midway] (end-edge-anc) {} (a-dmnd.north);
    \end{scope}
  \end{pgfonlayer}
  \draw[wilson line amplitude=4, wilson line, red]
  (a-dmnd.north) -- (a-dmnd.east) --
  (b-dmnd.west) -- (b-dmnd.north);

  \draw[->,dashed,thick] (start-edge-anc) to[bend left] node[midway,below] {\(U_{-}\)} (end-edge-anc);

  \begin{scope}[every node/.style={fill=white, rounded corners, inner sep=.1em, fill opacity=0.9, text opacity=1}]
    \node at ($(a-dmnd.north)!.5!(a-dmnd.east)$) {\(Z_a\)};
    \node at ($(b-dmnd.west)!.5!(b-dmnd.north)$) {\(X_b\)};
  \end{scope}

  \begin{scope}[every node/.style={majorana labelled vertex, minimum size=.8em}]
    \node at (a-dmnd.north) {1};
    \node at (b-dmnd.north) {2};
  \end{scope}

\end{tikzpicture}

%% file: figures/virtual-particle-final-state.tex
\begin{tikzpicture}[mjind/.style = {mj indicator, minimum size=.2cm}]
  \def\scale{1.7}
  
  \node[rectangle, inner xsep = \scale*2cm, inner ysep = \scale*1cm,overlay] (rct) at (0, 0) {};
  \begin{scope}[every node/.style={mjind}]
    \foreach \nn/\npos in
    {%
      nw/north west,%
      sw/south west,%
      ne/north east,%
      se/south east,%
      nc/north,%
      sc/south%
    } {
      \coordinate (\nn) at (rct.\npos) {};
    }
    \coordinate (marker-left) at ($(rct.north west)!.5!(rct.north)$);
    \coordinate (marker-right) at ($(rct.north)!.5!(rct.north east)$);
    \coordinate (cw) at (rct.west -| marker-left) {};

    \node[stroke=dotted,fill=none,dashed,thick,minimum size=.4cm] (ce) at (rct.west -| marker-right) {};
  \end{scope}

  \def\minpad{.5em}
  \clip[use as bounding box] ($(sc)+(-1,-\minpad)$) -- ($(nc)+(-1,\minpad)$) -- ($(ne)+(\minpad,\minpad)$) -- ($(se) + (\minpad,-\minpad)$) -- cycle;

  \foreach \nn in {nw,sw,ne,se,sc} {
    \node[mjind] (\nn) at (\nn) {};
  }

  \node[mjind] at (ce) {};

  \def\npair{3}
  \foreach \px [parse=true] in {1,...,2*\npair} {
    \pgfmathsetmacro{\fpos}{\px/(2*\npair)}
    \coordinate (p-\px) at ($(nc)!\fpos!(ce)$);
  }

  \begin{scope}[every node/.style={font=\LARGE}]
    \node at ($(p-2)!.5!(p-3)$) {\(\varepsilon\)};
    \node at ($(p-4)!.5!(p-5)$) {\(\varepsilon\)};
  \end{scope}
  \begin{scope}[every path/.style={line width=.07cm}]
    \draw (nw) -- (ne);
    \draw (sw) -- (sc);

    \draw (nw) -- (sw);
    \draw (sc) -- (se);
    \draw (ne) -- (se);
  \end{scope}

\end{tikzpicture}

%% file: figures/virtual-particle-move.tex
\begin{tikzpicture}[mjind/.style = {mj indicator, minimum size=.2cm}]
  \def\scale{1.7}

  \pgfdeclarelayer{wilsonlines}
  \pgfsetlayers{wilsonlines,main}

  \node[rectangle, inner xsep = \scale*2cm, inner ysep = \scale*1cm,overlay] (rct) at (0, 0) {};
  \begin{scope}[every node/.style={mjind}]
    \foreach \nn/\npos in
    {%
      nw/north west,%
      sw/south west,%
      ne/north east,%
      se/south east,%
      nc/north,%
      sc/south%
    } {
      \coordinate (\nn) at (rct.\npos) {};
    }
    \coordinate (marker-left) at ($(rct.north west)!.5!(rct.north)$);
    \coordinate (marker-right) at ($(rct.north)!.5!(rct.north east)$);
    \coordinate (cw) at (rct.west -| marker-left) {};

    \node[stroke=dotted,fill=none,dashed,thick,minimum size=.4cm] (ce) at (rct.west -| marker-right) {};
  \end{scope}

  \def\minpad{.5em}
  \clip[use as bounding box] ($(sc)+(-1,-\minpad)$) -- ($(nc)+(-1,\minpad)$) -- ($(ne)+(\minpad,\minpad)$) -- ($(se) + (\minpad,-\minpad)$) -- cycle;

  \foreach \nn in {nw,sw,ne,se,nc,sc} {
    \node[mjind] (\nn) at (\nn) {};
  }

  \def\npair{3}
  \foreach \px [parse=true] in {1,...,2*\npair} {
    \pgfmathsetmacro{\fpos}{\px/(2*\npair)}
    \node[mjind,opacity=.4] (p-\px) at ($(nc)!\fpos!(ce)$) {};
  }
  \begin{pgfonlayer}{wilsonlines}
    \foreach \px [parse=true] in {1,3,...,2*\npair} {
      \pgfmathsetmacro{\pxx}{\px+1}
      \draw[wilson line] (p-\px) -- (p-\pxx);
    }
    \begin{scope}[every path/.style={wilson line, red}]
      \foreach \px [parse=true] in {3,5,...,2*\npair-1} {
        \pgfmathsetmacro{\pxx}{\px-1}
        \draw (p-\pxx) -- (p-\px);
      }
      \draw (nc) -- (p-1);
    \end{scope}
  \end{pgfonlayer}
  \begin{scope}[every path/.style={line width=.07cm}]
    \draw (nw) -- (nc);
    \draw (sw) -- (sc);

    \draw (nw) -- (sw);
    \draw (nc) -- (ne);
    \draw (sc) -- (se);
    \draw (ne) -- (se);
  \end{scope}

\end{tikzpicture}

%% file: figures/mj-4q-stab.tex
\begin{tikzpicture}
  \pgfdeclarelayer{stabedges}
  \pgfdeclarelayer{mjdiamonds}
  \pgfsetlayers{stabedges,mjdiamonds,main}
  \tikzset{mj diamond size=.7em}
  \def\rectheight{2.5em}
  \node[overlay,rectangle, inner xsep={\rectheight}, inner ysep=\rectheight] (rct) at (0, 0) {};
  \begin{pgfonlayer}{mjdiamonds}
    \foreach \name/\loc/\rot in
    {
      nw/north west/90,
      sw/south west/0,
      se/south east/90,
      ne/north east/0
    }
    {
      \pic[name=qb-\name,rotate=\rot,transform shape] at (rct.\loc) {flat pf diamond};
    }
    \pic[name=qb-a] at (rct) {flat pf diamond};
  \end{pgfonlayer}
  \begin{pgfonlayer}{stabedges}
    \begin{scope}[every path/.style=stabilizer edge]
      \draw (qb-nw-dmnd.west) -- (qb-sw-dmnd.north);
      \draw (qb-sw-dmnd.east) -- (qb-se-dmnd.north);
      \draw (qb-se-dmnd.east) -- (qb-ne-dmnd.south);
      \draw (qb-ne-dmnd.west) -- (qb-nw-dmnd.south);

      \draw (qb-a-dmnd.east) to[bend right] (qb-a-dmnd.north);
      \draw (qb-a-dmnd.west) to[bend right] (qb-a-dmnd.south);
    \end{scope}
  \end{pgfonlayer}

  \begin{scope}
    \gatearrow{qb-ne-dmnd}{mjz}{qb-a-dmnd}{mjx}{1};
    \gatearrow{qb-nw-dmnd}{mjz}{qb-a-dmnd}{mjx}{2};
    \gatearrow{qb-se-dmnd}{mjz}{qb-a-dmnd}{mjx}{3};
    \gatearrow{qb-sw-dmnd}{mjz}{qb-a-dmnd}{mjx}{4};
  \end{scope}
\end{tikzpicture}

%% file: figures/mj-5q-stab.tex
\begin{tikzpicture}
  \pgfdeclarelayer{stabedges}
  \pgfdeclarelayer{mjdiamonds}
  \pgfsetlayers{stabedges,mjdiamonds,main}
  \tikzset{mj diamond size=.7em}
  \def\rectheight{2.5em}
  \node[overlay,rectangle, inner xsep={2*\rectheight}, inner ysep=\rectheight] (rct) at (0, 0) {};
  \begin{pgfonlayer}{mjdiamonds}
    \foreach \name/\loc in
    {nw/north west,
      sw/south west,
      ne/north east,
      n/north}
    {
      \pic[name=qb-\name] at (rct.\loc) {flat pf diamond};
    }
    \def\noff{0}
    \pic[name=qb-s,transform shape] at ($(rct.south)+(0,\noff)$) {flat pf diamond};
    \coordinate (ctopleft) at ($(qb-nw-dmnd)!.5!(qb-n-dmnd)$);
    \coordinate (ctopright) at ($(qb-n-dmnd)!.5!(qb-ne-dmnd)$);
    \coordinate (cleft) at ($(qb-sw-dmnd)!.5!(qb-nw-dmnd)$);
    \pic[name=qb-a-left] at (cleft -| ctopleft) {flat pf diamond};
    \pic[name=qb-a-right] at (cleft -| ctopright) {flat pf diamond};
  \end{pgfonlayer}
  \begin{pgfonlayer}{stabedges}
    \begin{scope}[every path/.style=stabilizer edge]
      \draw (qb-nw-dmnd.east) -- (qb-n-dmnd.west);
      \draw (qb-nw-dmnd.south) -- (qb-sw-dmnd.north);
      \draw (qb-sw-dmnd.east) -- (qb-s-dmnd.west);
      \draw (qb-s-dmnd.north) to[bend right, looseness=1.3] (qb-ne-dmnd.south);
      \draw (qb-ne-dmnd.west) -- (qb-n-dmnd.east);

      \draw (qb-a-left-dmnd.east) to[bend right] (qb-a-left-dmnd.north);
      \draw (qb-a-left-dmnd.west) to[bend right] (qb-a-left-dmnd.south);

      \draw (qb-a-right-dmnd.east) to[bend right] (qb-a-right-dmnd.north);
      \draw (qb-a-right-dmnd.west) to[bend right] (qb-a-right-dmnd.south);
    \end{scope}
  \end{pgfonlayer}

  \begin{scope}[gate indicator/.style={
      line width=0.04em,
      rectangle, inner xsep=1.25em, inner ysep=.9em, draw=black, sloped, rounded corners}]
    \gatearrow{qb-n-dmnd}{mjy}{qb-a-left-dmnd}{mjx}{1};
    \gatearrow{qb-nw-dmnd}{mjx}{qb-a-left-dmnd}{mjx}{2};
    \gatearrow{qb-s-dmnd}{mjx}{qb-a-left-dmnd}{mjx}{3};
    \gatearrow{qb-sw-dmnd}{mjz}{qb-a-left-dmnd}{mjx}{4};

    \gatearrow{qb-ne-dmnd}{mjz}{qb-a-right-dmnd}{mjx}{1,5};

    \path (qb-a-right-dmnd) -- node[midway, gate indicator, red] {} (qb-s-dmnd);
    \gatearrow{qb-a-right-dmnd}{mjz}{qb-s-dmnd}{mjz}{2,4};
  \end{scope}
\end{tikzpicture}

%% file: figures/mj-6q-stab.tex
\begin{tikzpicture}
  \pgfdeclarelayer{stabedges}
  \pgfdeclarelayer{mjdiamonds}
  \pgfsetlayers{stabedges,mjdiamonds,main}
  \tikzset{mj diamond size=.7em}
  \def\rectheight{2.5em}
  \node[overlay,rectangle, inner xsep={2*\rectheight}, inner ysep=\rectheight] (rct) at (0, 0) {};
  \begin{pgfonlayer}{mjdiamonds}
    \foreach \name/\loc in
    {nw/north west,
      sw/south west,
      ne/north east,
      n/north,
      s/south,
      se/south east}
    {
      \pic[name=qb-\name] at (rct.\loc) {flat pf diamond};
    }
    \coordinate (ctopleft) at ($(qb-nw-dmnd)!.5!(qb-n-dmnd)$);
    \coordinate (ctopright) at ($(qb-n-dmnd)!.5!(qb-ne-dmnd)$);
    \coordinate (cleft) at ($(qb-sw-dmnd)!.5!(qb-nw-dmnd)$);
    \pic[name=qb-a-left] at (cleft -| ctopleft) {flat pf diamond};
    \pic[name=qb-a-right] at (cleft -| ctopright) {flat pf diamond};
  \end{pgfonlayer}
  \begin{pgfonlayer}{stabedges}
    \begin{scope}[every path/.style=stabilizer edge]
      \draw (qb-nw-dmnd.east) -- (qb-n-dmnd.west);
      \draw (qb-nw-dmnd.south) -- (qb-sw-dmnd.north);
      \draw (qb-sw-dmnd.east) -- (qb-s-dmnd.west);
      \draw (qb-s-dmnd.north) -- (qb-se-dmnd.west);
      \draw (qb-ne-dmnd.west) -- (qb-n-dmnd.east);
      \draw (qb-se-dmnd.north) -- (qb-ne-dmnd.south);

      \draw (qb-a-left-dmnd.east) to[bend right] (qb-a-left-dmnd.north);
      \draw (qb-a-left-dmnd.west) to[bend right] (qb-a-left-dmnd.south);

      \draw (qb-a-right-dmnd.east) to[bend right] (qb-a-right-dmnd.north);
      \draw (qb-a-right-dmnd.west) to[bend right] (qb-a-right-dmnd.south);
    \end{scope}
  \end{pgfonlayer}

  \begin{scope}[gate indicator/.style={
      line width=0.04em,
      rectangle, inner xsep=1.25em, inner ysep=.9em, draw=black, sloped, rounded corners}]
    \gatearrow{qb-n-dmnd}{mjy}{qb-a-left-dmnd}{mjx}{1};
    \gatearrow{qb-nw-dmnd}{mjx}{qb-a-left-dmnd}{mjx}{2};
    \gatearrow{qb-s-dmnd}{mjx}{qb-a-left-dmnd}{mjx}{4};
    \gatearrow{qb-sw-dmnd}{mjz}{qb-a-left-dmnd}{mjx}{5};
    \gatearrow{qb-se-dmnd}{mjx}{qb-a-right-dmnd}{mjx}{2,6};

    \gatearrow{qb-ne-dmnd}{mjz}{qb-a-right-dmnd}{mjx}{1,7};

    \path (qb-a-right-dmnd) -- node[midway, gate indicator, red] {} (qb-s-dmnd);
    \gatearrow{qb-a-right-dmnd}{mjz}{qb-s-dmnd}{mjz}{3,5};
  \end{scope}
\end{tikzpicture}

%% file: figures/fermion-error-sc3.tex
\begin{tikzpicture}
  \pgfdeclarelayer{stabedges}
  \pgfdeclarelayer{mjdiamonds}
  \pgfsetlayers{stabedges,mjdiamonds,main}
  \tikzset{mj diamond size=.9em}

  \def\xdist{3}
  \def\ydist{3}

  \begin{pgfonlayer}{mjdiamonds}
    \begin{scope}[scale=1]
      \foreach \x [parse=true] in {0,...,\xdist-1} {
        \foreach \y [parse=true] in {0,...,\ydist-1} {
          \pgfmathtruncatemacro{\xc}{2*\x + 1}
          \pgfmathtruncatemacro{\yc}{2*\y + 1}
          \pic[rotate=45,transform shape] (qb-\xc-\yc) at (\xc,\yc) {flat pf diamond};
        }
      }
      \foreach \x [parse=true] in {1,...,\xdist-1} {
        \foreach \y [parse=true] in {1,...,\ydist-1} {
          \pgfmathtruncatemacro{\xc}{2*\x}
          \pgfmathtruncatemacro{\yc}{2*\y}
          \pic[rotate=45,transform shape] (qb-\xc-\yc) at (\xc,\yc) {flat pf diamond};
        }
      }
    \end{scope}
  \end{pgfonlayer}
  \begin{pgfonlayer}{stabedges}
    \begin{scope}[every path/.style={stabilizer edge}]
      \foreach \x [parse=true] in {0,...,\xdist-2} {
        \pgfmathtruncatemacro{\xc}{2*\x+1}
        \pgfmathtruncatemacro{\xco}{\xc+2}
        \pgfmathtruncatemacro{\yc}{2*\ydist-1}
        \draw (qb-\xc-1-dmnd.south) -- (qb-\xco-1-dmnd.west);
        \draw (qb-\xc-\yc-dmnd.east) -- (qb-\xco-\yc-dmnd.north);
      }
      \foreach \y [parse=true] in {0,...,\ydist-2} {
        \pgfmathtruncatemacro{\yc}{2*\y+1}
        \pgfmathtruncatemacro{\yco}{\yc+2}
        \pgfmathtruncatemacro{\xc}{2*\xdist-1}
        \draw (qb-1-\yc-dmnd.north) -- (qb-1-\yco-dmnd.west);
        \draw (qb-\xc-\yc-dmnd.east) -- (qb-\xc-\yco-dmnd.south);
      }
      \foreach \x [parse=true] in {1,...,\xdist-1} {
        \foreach \y [parse=true] in {1,...,\ydist-1} {
          \pgfmathtruncatemacro{\xa}{2*\x}
          \pgfmathtruncatemacro{\ya}{2*\y}
          \foreach \dx/\dy/\anchor/\aanchor in {
            -1/-1/east/west,
            1/-1/north/south,
            1/1/west/east,
            -1/1/south/north} {
            \pgfmathtruncatemacro{\xd}{\xa+\dx}
            \pgfmathtruncatemacro{\yd}{\ya+\dy}
            \draw (qb-\xa-\ya-dmnd.\aanchor) -- (qb-\xd-\yd-dmnd.\anchor);
          }
        }
      }
    \end{scope}
\end{pgfonlayer}

\begin{scope}[gate arrow/.style={
    line width=.3em,
    arrows={Circle[length=.5em] - Circle[length=.5em]},
    mjy,opacity=0.8
  }]
  \def\drawga#1#2{%
    \draw[gate arrow] (#1.center) -- (#2.center);
  }
  \drawga{qb-1-1-dmnd}{qb-2-2-dmnd}

  \drawga{qb-2-4-dmnd}{qb-3-5-dmnd}
  \drawga{qb-3-3-dmnd}{qb-4-4-dmnd}
  \drawga{qb-4-2-dmnd}{qb-5-3-dmnd}
\end{scope}
\end{tikzpicture}

%% file: figures/lattice-surgery.tex
\begin{tikzpicture}[rotate=-45,transform shape,mj diamond size=8]
  \pgfdeclarelayer{stabedges}
  \pgfdeclarelayer{mjdiamonds}
  \pgfsetlayers{stabedges,mjdiamonds,main}
  \def\scxdist{3}
  \def\scydist{3}
  \begin{scope}[scale=.8]
    \begin{pgfonlayer}{mjdiamonds}
      \foreach \x[parse=true] in {1,...,2*\scxdist} {
        \foreach \y in {1,...,\scydist} {
          \pgfmathtruncatemacro{\x}{\x}
          \pgfmathtruncatemacro{\y}{\y}
          \pic[name=qb-\x-\y] at (\x,\y) {flat pf diamond};
        }
      }
      \foreach \x[parse=true] in {\scxdist+1,...,2*\scxdist} {
        \foreach \y[parse=true] in {\scydist+1,...,2*\scydist}
        \pgfmathtruncatemacro{\x}{\x}
        \pgfmathtruncatemacro{\y}{\y}
        \pic[name=qb-\x-\y] at (\x,\y) {flat pf diamond};
      }
    \end{pgfonlayer}
  \end{scope}
  \begin{pgfonlayer}{stabedges}
    \begin{scope}[every path/.style=stabilizer edge]
      \foreach \x in {1,...,\scxdist} {
        \foreach \y in {1,...,\scydist} {
          \ifnum \x<\scxdist
            \pgfmathtruncatemacro{\nextx}{\x+1}
            \draw (qb-\x-\y-dmnd.east) -- (qb-\nextx-\y-dmnd.west);
          \fi
          \ifnum \y<\scydist
            \pgfmathtruncatemacro{\nexty}{\y+1}
            \draw (qb-\x-\y-dmnd.north) -- (qb-\x-\nexty-dmnd.south);
          \fi
        }
      }
      \pgfmathtruncatemacro{\mxmax}{2*\scxdist}
      \pgfmathtruncatemacro{\mxmin}{\scxdist+1}
      \foreach \x in {\mxmin,...,\mxmax} {
        \foreach \y in {1,...,\scydist} {
          \ifnum \x<\mxmax
            \pgfmathtruncatemacro{\nextx}{\x+1}
            \draw (qb-\x-\y-dmnd.east) -- (qb-\nextx-\y-dmnd.west);
          \fi
          \ifnum \y<\scydist
            \pgfmathtruncatemacro{\nexty}{\y+1}
            \draw (qb-\x-\y-dmnd.north) -- (qb-\x-\nexty-dmnd.south);
          \fi
        }
      }
      \pgfmathtruncatemacro{\mymax}{2*\scydist}
      \pgfmathtruncatemacro{\mymin}{\scydist+1}
      \foreach \x in {\mxmin,...,\mxmax} {
        \foreach \y in {\mymin,...,\mymax} {
          \ifnum \x<\mxmax
            \pgfmathtruncatemacro{\nextx}{\x+1}
            \draw (qb-\x-\y-dmnd.east) -- (qb-\nextx-\y-dmnd.west);
          \fi
          \ifnum \y<\mymax
            \pgfmathtruncatemacro{\nexty}{\y+1}
            \draw (qb-\x-\y-dmnd.north) -- (qb-\x-\nexty-dmnd.south);
          \fi
        }
      }
      \foreach \x in {1,...,\scxdist} {
        \pgfmathtruncatemacro{\nextx}{\x+1}
        \ifnum\nextx>\scxdist\else
          \ifodd\x
            \draw (qb-\x-\scydist-dmnd.north) to[bend left]
            (qb-\nextx-\scydist-dmnd.north);
          \else
            \draw (qb-\x-1-dmnd.south) to[bend right]
            (qb-\nextx-1-dmnd.south);
          \fi
        \fi
      }
      \foreach \y in {1,...,\scydist} {
        \pgfmathtruncatemacro{\nexty}{\y+1}
        \ifnum\nexty>\scydist\else
          \ifodd\y
            \draw (qb-1-\y-dmnd.west) to[bend left]
            (qb-1-\nexty-dmnd.west);
          \else
            \draw (qb-\scxdist-\y-dmnd.east) to[bend right]
            (qb-\scxdist-\nexty-dmnd.east);
          \fi
        \fi
      }

      \foreach \x in {\mxmin,...,\mxmax} {
        \pgfmathtruncatemacro{\nextx}{\x+1}
        \ifnum\nextx>\mxmax\else
          \ifodd\nextx
            \draw (qb-\x-\scydist-dmnd.north) to[bend left]
            (qb-\nextx-\scydist-dmnd.north);
          \else
            \draw (qb-\x-1-dmnd.south) to[bend right]
            (qb-\nextx-1-dmnd.south);
          \fi
        \fi
      }
      \foreach \y in {1,...,\scydist} {
        \pgfmathtruncatemacro{\nexty}{\y+1}
        \ifnum\nexty>\scydist\else
          \ifodd\y
            \draw (qb-\mxmin-\y-dmnd.west) to[bend left]
            (qb-\mxmin-\nexty-dmnd.west);
          \else
            \draw (qb-\mxmax-\y-dmnd.east) to[bend right]
            (qb-\mxmax-\nexty-dmnd.east);
          \fi
        \fi
      }

      \foreach \x in {\mxmin,...,\mxmax} {
        \pgfmathtruncatemacro{\nextx}{\x+1}
        \ifnum\nextx>\mxmax\else
          \ifodd\nextx
            \draw (qb-\x-\mymax-dmnd.north) to[bend left]
            (qb-\nextx-\mymax-dmnd.north);
          \else
            \draw (qb-\x-\mymin-dmnd.south) to[bend right]
            (qb-\nextx-\mymin-dmnd.south);
          \fi
        \fi
      }
      \foreach \y in {\mymin,...,\mymax} {
        \pgfmathtruncatemacro{\nexty}{\y+1}
        \ifnum\nexty>\mymax\else
          \ifodd\nexty
            \draw (qb-\mxmin-\y-dmnd.west) to[bend left]
            (qb-\mxmin-\nexty-dmnd.west);
          \else
            \draw (qb-\mxmax-\y-dmnd.east) to[bend right]
            (qb-\mxmax-\nexty-dmnd.east);
          \fi
        \fi
      }
    \end{scope}
  \end{pgfonlayer}
\end{tikzpicture}

%% file: SupplementaryMaterial.tex
\section{Basics of Majorana code theory}

The purpose of this section is to summarize notions from Majorana graph theory that are not standard knowledge in surface code error correction theory. For detailed discussion of theory~\cite{LENSKY2023169286}.

\subsection{Majorana fermions, fermions and Pauli matrices}

Consider a qubit, which is nothing but a two level system. All qubit operations can be described in the language of fermion creation/annihilation operators $c^\dagger, c$, that satisfy, 
\begin{gather}
    c^\dagger c + c c^\dagger = 1, \qquad c^2 =0.
\end{gather} 
The same description is also realized in terms of a pair of Majorana fermion operators, $\alpha_1, \alpha_2$, where $\alpha_j=\alpha^\dagger_j$, and
\begin{gather}
\alpha_j \alpha_k + \alpha_k \alpha_j = 2\delta_{jk},
\end{gather}
then fermion creation/annihilation operators can be introduced as,
\begin{gather}
c^\dagger = \frac{1}{2} (\alpha_2 + i \alpha_1), \qquad c = \frac{1}{2}(\alpha_2 -i\alpha_1). \label{eq:fermion-c-cdag}
\end{gather}
A Majorana bilinear measures fermion parity $i\alpha_2 \alpha_1 = 1-2 c^\dagger c = (-1)^{c^\dagger c}$. Remarkably, $\alpha_1$ and $\alpha_2$ can be spatially separated. In this way a pair of Majorana fermions encode one qubit non-locally. 

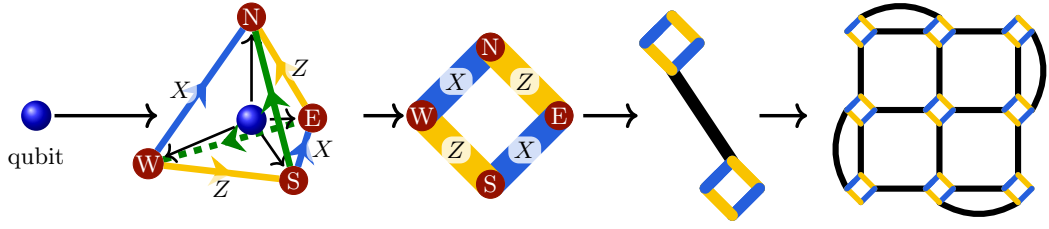
\begin{figure*}[t]
  \centering
  \input{figures/gauge-theory-outline-labelled.tex}
  \caption{Construction of the Majorana graph of a distance 3 surface code. From left to right: (i) Mapping a qubit into four Majorana fermions, red dots labeled with cardinal directions N, E, S, W; Bilinears of Majorana fermions corresponding corresponding to single qubit Pauli operators Eq.~(\ref{eq:XYZviaMJ}) are shown as colored lines. Arrows show the order of Majorana fermions in the definition of the Pauli. (ii) Diamond graphical notation for the qubit. (iii) A pair of qubits connected by a Majorana dimer. (iv) Majorana graph for distance 3 code. }
  \label{fig:mj-diamond-NESW}
\end{figure*}

For a single qubit the algebra Pauli operators, $X, Y, Z$ matches Majorana fermions,
\begin{gather}
X = \alpha_1, \; Y= \alpha_2, \; Z= -i \alpha_1 \alpha_2, \label{eq:1QPauliVMajorana}
\end{gather}
and it is tempting to say that the spin problem is reduced to non-interacting fermions.
However, Pauli operators on different qubits commute whereas Majorana operators anti-commute. This difficulty can be effectively bypassed by introduction the Jordan-Wigner string~\cite{JordanWigner}. In 2D Jordan-Wigner strings become too cumbersome to be useful. We adopt representation of spin operators fundamentally different from Eq.~(\ref{eq:1QPauliVMajorana}). Namely, we first double the Hilbert space by considering four Majorana fermions, $\alpha_N, \alpha_E, \alpha_S, \alpha_W$, labeled North, East, South, West, respectively, see Fig.~\ref{fig:mj-diamond-NESW}. The identification with Pauli operators shown pictorially in Fig.~\ref{fig:mj-diamond-NESW} is chosen to satisfy Pauli algebra,
\begin{align}
X &\in \{i \alpha_N \alpha_W, - i \alpha_S\alpha_E\}, \nonumber \\
Z &\in \{i \alpha_N \alpha_S, -i \alpha_E \alpha_W\},  ~\label{eq:XYZviaMJ} \\
Y &\in \{ i \alpha_S \alpha_W, - i \alpha_N \alpha_E \}. \nonumber
\end{align}
We mean that if any representatives for \(X,Y,Z\) are chosen from the sets
above, they will form a Pauli algebra.

Gluing identical representations of Pauli operators in terms of Majorana fermions requires the wave function to satisfy the constraint,
\begin{gather}
 \alpha_N  \alpha_E \alpha_S \alpha_W \ket{\psi} = \ket{\psi}.\label{eq:GaugeOperatorGamma}
\end{gather}
that reduces the Hilbert space of to that of a single qubit.

\subsection{Majorana vs Pauli stabilizer code}

Consider a stabilizer of a standard surface code, Fig.~\ref{fig:mj-diamond-NESW}, that has a form in a symmetric basis,
\begin{gather}
B = Z_1 X_2 Z_3 X_4.
\end{gather}
Substituting Majorana representation from eq.~(\ref{eq:XYZviaMJ}) and regrouping Majorana operators we obtain,
\begin{gather}
B = \tau_{12} \tau_{23} \tau_{34} \tau_{41}, \label{eq:B-vs-tau}
\end{gather}
where we introduced link parity operators,
\begin{gather}
\tau_{jk} = i \alpha_j \alpha_k, \label{eq:link-parity}
\end{gather}
connecting qubits $j$ and $k$. Here the Majorana fermion cardinal direction index, North, East, South, West, is augmented with the qubit index, such that for example, $\tau_{14} = i \alpha_{1S} \alpha_{4N}$. The order of Majorana fermions in the definition of $\tau_{jk}$ is important to determine the correct sign. The sign is prescribed by a Kasteleyn field, $\mathbb{Z}_2^{K}$, formally described in Ref.~\cite{LENSKY2023169286}. For the purposes of this work such details are not necessary we refer interested readers to Ref.~\cite{LENSKY2023169286}. We now have a procedure that maps between a Majorana graph and stabilizer code: each qubit is replaced with a diamond representing four Majorana fermions, then stabilizers are formed by loops of link parity operators. The constraints Eq.~(\ref{eq:GaugeOperatorGamma}) apply at every qubit and act similarly to a Gauss law. The stabilizer condition 
\begin{gather}
B(P)\ket{\psi}=\ket{\psi}, \label{eq:stab-cond}
\end{gather}
resembles Eq.~(\ref{eq:GaugeOperatorGamma}). However, unlike the latter which is absolute the former describes the protected computational space of the code and may be violated. The constraints in Eq.~(\ref{eq:stab-cond}) correspond to the absence of flux, or flatness condition. The stabilizer $B$ is a product of link parity operators, Eq.~(\ref{eq:B-vs-tau}), that form a loop and is an analog of a loop integral of a gauge field where the emergent gauge field operators $\ZIIS$ are link parity operators Eq.~(\ref{eq:link-parity}).

Surface code maps onto a set of Majorana dimers (links). The bulk of the standard surface code $C$ consists of dimerized Majoranas which are essentially frozen by the conditions Eq.~(\ref{eq:GaugeOperatorGamma}) for each qubit and Eq.~(\ref{eq:stab-cond}) for each stabilizer. Note that Majorana fermions on corner qubits have only three links rather than four, see Fig.~\ref{fig:mj-diamond-NESW} right. There are four such undimerized Majorana fermions in total. However, they satisfy even parity constraint~\cref{eq:GaugeOperatorGamma} and therefore surface code stores only one logical qubit. Logical operators are determined analogously to~\cref{eq:XYZviaMJ} with the difference that the Majorana fermions correspond to different corner qubits.  

Removing a single link from a code $C$ creates a new code $C^\prime$ that has a pair of undimerized Majoranas $\sigma_{1,2}$, see Fig. ~\ref{fig:removed-link-d6}. In this configuration the two stabilizers $B_1$, $B_2$ of $C$ that shared the removed edge are replaced with a single stabilizer equal to their product $B_1B_2$ in $C^\prime$. As a result the computational space is doubled: in addition to vacuum, $B_1 \ket{\psi}=B_2\ket{\psi} = \ket{\psi}$, there is a fermion $\varepsilon$ state characterized by $B_1 \ket{\psi}=B_2\ket{\psi} = -\ket{\psi}$. The latter is the familiar $\varepsilon$ excitation from the theory of surface code corresponding to two adjacent stabilizer violations~\cite{dennisTopologicalQuantumMemory2002}. The relationship between the two codes, $C_0$ and $C^\prime$, is described by the fusion rule in Eq.~(\ref{eq:mj-fusion-rule}),
\begin{gather}
\sigma \times \sigma = 1 + \varepsilon.
\end{gather}
Indeed, removing a link creates a pair of $\sigma$ regardless of the original state being vacuum or a single fermion $\varepsilon$. Conversely, measuring two stabilizers $B_1$ and $B_2$ instead of one $B_1B_2$, i.e. restoring the dimer link results in either vacuum or a fermion, depending on the state of the pair of $\sigma$. The Majorana bilinear operator corresponding to the link $i\sigma_{1}\sigma_{2} = (-1)^{\varepsilon^\dagger \varepsilon}$ measures fermion density by construction. It is natural to ask which operators flip the parity, namely creation and annihilation operators, that are related to Majorana operators by Eq.~(\ref{eq:fermion-c-cdag}), $\varepsilon = \tfrac{1}{2} (\sigma_2 - i\sigma_1)$. $\varepsilon$ corresponding to different dimer links anti-commute.  The same fact is encoded as a fusion rule Eq.~(\ref{eq:mj-fusion-rule}). 

\begin{widetext}

\begin{figure}[h!]
  \centering
  \input{figures/RemovedLink-fig}
  \caption{Left: A region of the Majorana graph of a surface code. Right: The same as left with a single Majorana dimer removed. }
  \label{fig:removed-link-d6}
\end{figure}
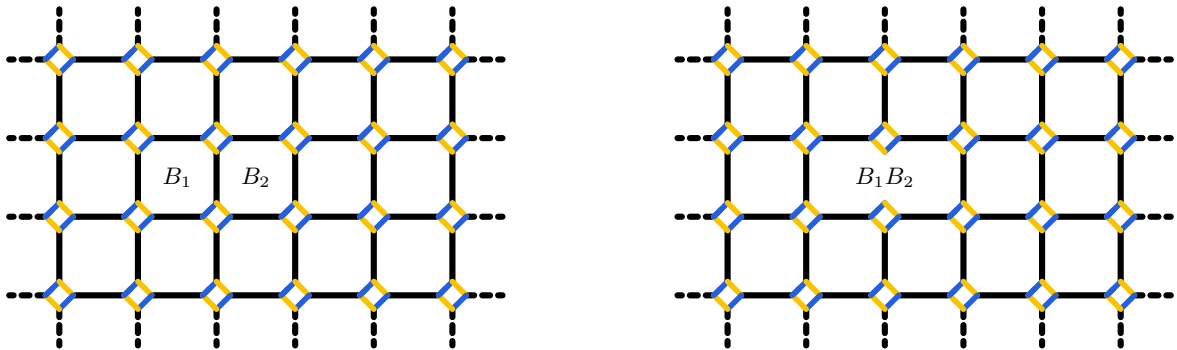

\end{widetext}

Removing a link from a surface code creates a pair of lattice dislocation defects: vertices with degree three on the graph whereas in the bulk of the surface code all graph vertices have degree four.  These defects can be separated spatially without creating additional degree three defects by creating a dislocation line between them, see Fig.~\ref{fig:3q-dense-mj-graph}. Because the Majorana description is exact on the lattice these degree three vertices host undimerized Majorana fermions, which are point-like particles on the graph. 
In Ref.~\cite{LENSKY2023169286} it was shown that Majorana fermions hosted by such degree three vertices in the surface code are bound to a flux $\mathbb{Z}_2^{(K)}$. Therefore the Majorana fermions behave as Ising non-Abelian anyons. Spatially separated defects therefore form logical qubits.

The fusion rule Eq.~(\ref{eq:mj-fusion-rule}) suggests that Majorana fermion is the fundamental degree of freedom in the surface code in the sense that all other excitations are emergent from them. Eq.~(\ref{eq:mj-fusion-rule}) describes the emergence of fermion $\varepsilon$. Two fermions sharing a stabilizer plaquette correspond to two stabilizer violations. Such stabilizer violations carry flux \(\ZIIS\) and therefore are Abelian anyons~\cite{dennisTopologicalQuantumMemory2002}. In this sense the latter particles are emergent from fermions $\varepsilon$. In absence of bulk Majorana fermions such stabilizer violations are called $e$ and $m$. However, in presence of undimerized Majorana fermions in the bulk particles $e$ and $m$ are no welllonger  defined: if $e$ or $m$ is braided around $\sigma$ the geometry of the lattice with a dislocation transforms $e$ into $m$ and vice versa. It is more convenient therefore to use equivalent and simpler description in terms of $\varepsilon$ fermions for the purpose of error analysis. 

One qubit Pauli errors are Majorana bilinears on a single qubit Eq.~(\ref{eq:XYZviaMJ}). It therefore creates pairs of fermions $\varepsilon$. Strings of fermions end with stabilizer violations. Such strings correspond to 't Hooft lines in the gauge field theory, introduced in the main text. Fermion parity requires that 't Hooft line crosses an even number of links. An 't Hooft line that is either closed or starts and ends at a boundary does not create stabilizer violations. An 't Hooft line that encloses an even number of undimerized Majorana fermions measures fermion parity. Such lines therefore correspond to undetectable errors and need to be accounted in the computation of code distance. The specific procedure for computing the distance is described in the main text.   

In Ref.~\cite{LENSKY2023169286} it was shown that an 't Hooft loop enclosing two undimerized Majorana fermions is equivalent to Wilson line: the minimal length path on the graph that connects Majorana fermions. In many geometries and specifically for bulk Majorana fermions far from boundaries Wilson lines correspond to the shortest undetectable error. This identification between Wilson and 't Hooft lines also provides an intuitive picture for undetectable error: it corresponds to a transport of a fermion $\varepsilon$ from one Majorana to another.   

\onecolumngrid

\section{Circuit slices}
\label{sec:circuit-slices}

We present the full measurement circuit for the most complicated code in the
virtual particle braid \cref{fig:unoptimized-stab-configs}; the others are
constructed similarly. This circuit both extracts syndromes, and measures the
fusion outcomes as in \cref{fig:mj-virtual-motion}. It consists of two
repeated cycles, which between them extract all the stabilizers and fusion
outcomes.

\begin{figure}[htbp]
  \centering
  \subfloat[Cycle 1.]{
    \includegraphics[width=0.9\textwidth]{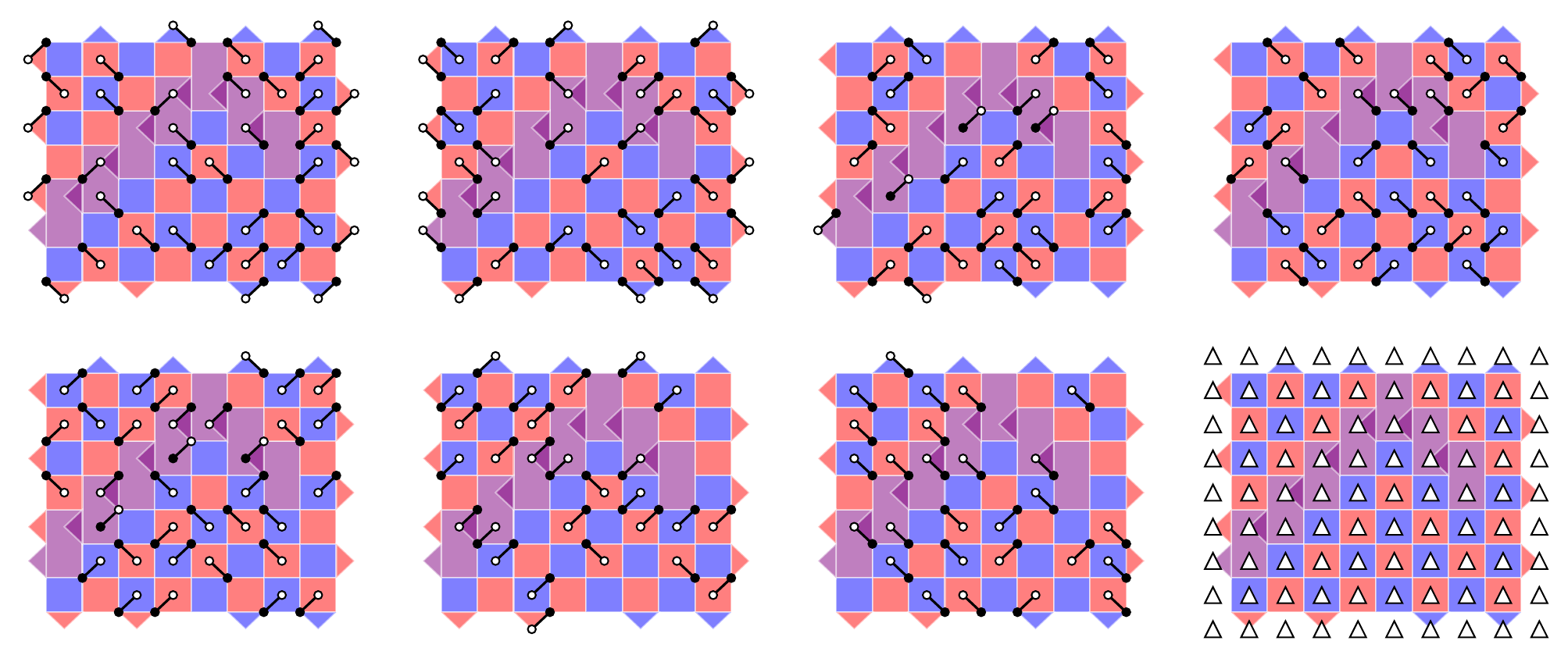}
  } \\
  \subfloat[Cycle 2.]{
    \includegraphics[width=0.9\textwidth]{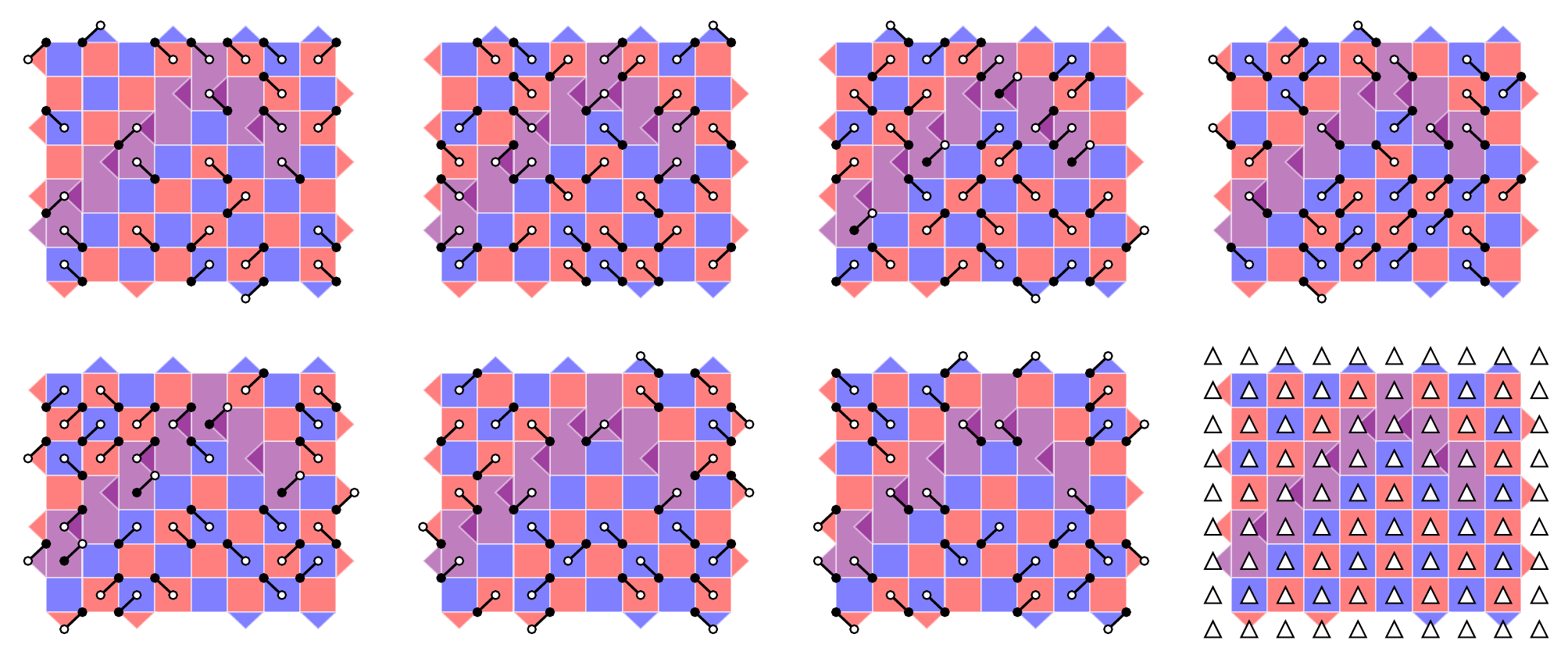}
  }
  \caption{Measurement circuits for the 5th stabilizer configuration of the
    braid shown in \cref{fig:unoptimized-stab-configs}. The single-qubit
    gates depend on the stabilizer basis, but are chosen precisely as in
    \cref{fig:general-stab-mu-circuits}. The only remaining choice is the
    particular pattern and ordering of the 2-qubit gates, which is shown
    here. Measurement and subsequent resets on ancilla qubits are shown as
    white triangles.}
  \label{fig:measurement-circuits}
\end{figure}


%% file: figures/gauge-theory-outline-labelled.tex
  \begin{tikzpicture}[scale=0.8,
    transition-arrow/.style={->, line width=1.3pt},
    physical-qubit/.append style={minimum size=.4cm},
    majorana diamond edge/.append style={line width=2.5},
    plabel/.style={black, fill=white, fill opacity=.7, inner sep=2, rounded corners, text opacity=1}]
    \def\tetralen{1.7}
    \pgfmathsetmacro{\tetrang}{180-asin(2 * sqrt(2)/3)}
    \pgfmathsetmacro{\tetrarad}{sqrt(3/2)/2}
    \pgfmathsetmacro{\tetrarot}{45}
    \coordinate (qubit) at (0,0,0) {};
    \begin{scope}
      \begin{scope}[every node/.style={majorana vertex}, scale=\tetralen, shift=(qubit)]
        \foreach \pn/\lbl in {1/S,2/W,3/E} {
          \node[text=white] (mj-\pn) at ($({sin(\tetrang) * cos((\pn-1)*120 + \tetrarot)}, {cos(\tetrang)}, {sin(\tetrang) * sin((\pn-1)*120 + \tetrarot)})$) {\lbl};
        }
        \node[text=white] (mj-4) at (0, 1, 0) {N};
      \end{scope}
      \draw[majorana diamond edge, rev kasteleyn arrow, tau3link, dashed] (mj-2) -- (mj-3);
      \node[physical-qubit, minimum size=.4cm] (qubit) at (qubit) {};
      \begin{scope}[every path/.style={line width=1pt, ->, shorten >= 1pt, shorten <= 1pt}]
        \foreach \x in {1,...,4} {
          \draw (qubit) -- (mj-\x);
        }
      \end{scope}
      \begin{scope}[every path/.style={majorana diamond edge}, local bounding box=tetrabox]
        \draw[rev kasteleyn arrow, mjz] (mj-1) -- node[midway, plabel, below] {\(Z\)} (mj-2);
        \draw[kasteleyn arrow, mjz] (mj-3) -- node[midway, plabel, right] {\(Z\)} (mj-4);
        \draw[kasteleyn arrow, mjx] (mj-1) -- node[midway, plabel, right] {\(X\)} (mj-3);
        \draw[kasteleyn arrow, mjx] (mj-2) -- node[midway, plabel, left] {\(X\)} (mj-4);
        \draw[kasteleyn arrow, mjy] (mj-1) -- (mj-4);
      \end{scope}
    \end{scope}
    \node[physical-qubit] (initial-qubit) at ($(tetrabox.west) + (-2,0)$) {};
    \node[below=.3] at (initial-qubit) {qubit};
    \draw[transition-arrow, shorten <= 1] (initial-qubit) -- (tetrabox.west);
    \pic[mj diamond size=1.8cm] (diamond) at ($(tetrabox.east)+(2.5,0)$) {flat pf diamond};
    \begin{scope}[every node/.style={majorana labelled vertex}]
      \node at ($(diamond-dmnd.north)$) {N};
      \node at ($(diamond-dmnd.south)$) {S};
      \node at ($(diamond-dmnd.east)$) {E};
      \node at ($(diamond-dmnd.west)$) {W};
    \end{scope}
    \begin{scope}[every node/.style={plabel, fill opacity=0.8}]
      \node at ($(diamond-dmnd.east)!.5!(diamond-dmnd.north)$) {\(Z\)};
      \node at ($(diamond-dmnd.west)!.5!(diamond-dmnd.north)$) {\(X\)};
      \node at ($(diamond-dmnd.west)!.5!(diamond-dmnd.south)$) {\(Z\)};
      \node at ($(diamond-dmnd.south)!.5!(diamond-dmnd.east)$) {\(X\)};
    \end{scope}
    \draw[transition-arrow] ($(tetrabox.east)+(0.4, 0)$) -- ($(diamond-dmnd.west)+(-0.3,0)$);
    \begin{scope}[mj diamond size=20, local bounding box=single-edge-bb]
      \pic (d1) at ($(diamond-dmnd) + (3,1.2)$) {flat pf diamond};
      \pic (d2) at ($(diamond-dmnd) + (4,-1.2)$) {flat pf diamond};
      \draw[stabilizer edge] (d1-dmnd.south) -- (d2-dmnd.north);
      \pic (d1) at ($(diamond-dmnd) + (3,1.2)$) {flat pf diamond};
      \pic (d2) at ($(diamond-dmnd) + (4,-1.2)$) {flat pf diamond};
    \end{scope}
    \draw[transition-arrow] ($(diamond-dmnd.east) + (0.4,0)$) -- ++(.9, 0);

    \pgfdeclarelayer{stabedges}
    \pgfdeclarelayer{mjdiamonds}
    \pgfsetlayers{stabedges,mjdiamonds,main}
    \begin{scope}[xshift=27em,yshift=-7.5em, mj diamond size=10, local bounding box=surface-code-bb]
      \def\scxdist{3}
      \def\scydist{3}
      \begin{scope}[scale=1.3]
        \begin{pgfonlayer}{mjdiamonds}
          \foreach \x in {1,...,\scxdist} {
            \foreach \y in {1,...,\scydist} {
              \pic[name=qb-\x-\y] at (\x,\y) {flat pf diamond};
            }
          }
        \end{pgfonlayer}
      \end{scope}
      \begin{pgfonlayer}{stabedges}
        \begin{scope}[every path/.style=stabilizer edge]
          \foreach \x in {1,...,\scxdist} {
            \foreach \y in {1,...,\scydist} {
              \ifnum \x<\scxdist
                \pgfmathtruncatemacro{\nextx}{\x+1}
                \draw (qb-\x-\y-dmnd.east) -- (qb-\nextx-\y-dmnd.west);
              \fi
              \ifnum \y<\scydist
                \pgfmathtruncatemacro{\nexty}{\y+1}
                \draw (qb-\x-\y-dmnd.north) -- (qb-\x-\nexty-dmnd.south);
              \fi
            }
          }
          \foreach \x in {1,...,\scxdist} {
            \pgfmathtruncatemacro{\nextx}{\x+1}
            \ifnum\nextx>\scxdist\else
              \ifodd\x
                \draw (qb-\x-\scydist-dmnd.north) to[bend left]
                (qb-\nextx-\scydist-dmnd.north);
              \else
                \draw (qb-\x-1-dmnd.south) to[bend right]
                (qb-\nextx-1-dmnd.south);
              \fi
            \fi
          }
          \foreach \y in {1,...,\scydist} {
            \pgfmathtruncatemacro{\nexty}{\y+1}
            \ifnum\nexty>\scydist\else
              \ifodd\y
                \draw (qb-1-\y-dmnd.west) to[bend left]
                (qb-1-\nexty-dmnd.west);
              \else
                \draw (qb-\scxdist-\y-dmnd.east) to[bend right]
                (qb-\scxdist-\nexty-dmnd.east);
              \fi
            \fi
          }
        \end{scope}
      \end{pgfonlayer}
    \end{scope}

    \draw[transition-arrow, shorten >= 1em] (single-edge-bb.east) -- (single-edge-bb.east -| surface-code-bb.west);
  \end{tikzpicture}

%% file: figures/RemovedLink-fig.tex
  \begin{tikzpicture}[scale=0.8,
    transition-arrow/.style={->, line width=1.3pt},
    physical-qubit/.append style={minimum size=.4cm},
    majorana diamond edge/.append style={line width=2.5},
    plabel/.style={black, fill=white, fill opacity=.7, inner sep=2, rounded corners, text opacity=1}]
    
    \def\scxdist{6}
    \def\scydist{4}

    \pgfdeclarelayer{stabedges}
    \pgfdeclarelayer{mjdiamonds}
    \pgfsetlayers{stabedges,mjdiamonds,main}

    \begin{scope}[shift={(0,0)}, mj diamond size=10, local bounding box=surface-code-left]
      \begin{scope}[scale=1.3]
        \begin{pgfonlayer}{mjdiamonds}
          \foreach \x in {1,...,\scxdist} {
            \foreach \y in {1,...,\scydist} {
              \pic[name=qb1-\x-\y] at (\x,\y) {flat pf diamond};
            }
          }
        \end{pgfonlayer}
      \end{scope}
      \begin{pgfonlayer}{stabedges}
        \begin{scope}[every path/.style=stabilizer edge]
          \foreach \x in {1,...,\scxdist} {
            \foreach \y in {1,...,\scydist} {
              
              \ifnum \x<\scxdist
                \pgfmathtruncatemacro{\nextx}{\x+1}
                \draw (qb1-\x-\y-dmnd.east) -- (qb1-\nextx-\y-dmnd.west);
              \fi
              
              \ifnum \y<\scydist
                \pgfmathtruncatemacro{\nexty}{\y+1}
                \draw (qb1-\x-\y-dmnd.north) -- (qb1-\x-\nexty-dmnd.south);
              \fi
              
            }
          }
          
          \foreach \x in {1,...,\scxdist} {
            \draw[dashed] (qb1-\x-1-dmnd.south) -- +(0, -0.6);
            \draw[dashed] (qb1-\x-\scydist-dmnd.north) -- +(0, 0.6);
          }
          \foreach \y in {1,...,\scydist} {
            \draw[dashed] (qb1-1-\y-dmnd.west) -- +(-0.6, 0);
            \draw[dashed] (qb1-\scxdist-\y-dmnd.east) -- +(0.6, 0);
          }
        \end{scope}
      \end{pgfonlayer}

      \node[plabel] at ($(qb1-2-2-dmnd.center)!.5!(qb1-3-3-dmnd.center)$) {\(B_1\)};
      \node[plabel] at ($(qb1-3-2-dmnd.center)!.5!(qb1-4-3-dmnd.center)$) {\(B_2\)};
    \end{scope}

    \begin{scope}[xshift=34em, mj diamond size=10, local bounding box=surface-code-right]
      \begin{scope}[scale=1.3]
        \begin{pgfonlayer}{mjdiamonds}
          \foreach \x in {1,...,\scxdist} {
            \foreach \y in {1,...,\scydist} {
              \pic[name=qb2-\x-\y] at (\x,\y) {flat pf diamond};
            }
          }
        \end{pgfonlayer}
      \end{scope}
      \begin{pgfonlayer}{stabedges}
        \begin{scope}[every path/.style=stabilizer edge]
          \foreach \x in {1,...,\scxdist} {
            \foreach \y in {1,...,\scydist} {
                
              \ifnum \x<\scxdist
                \pgfmathtruncatemacro{\nextx}{\x+1}
                \draw (qb2-\x-\y-dmnd.east) -- (qb2-\nextx-\y-dmnd.west);
              \fi
              
              \ifnum \y<\scydist
                \pgfmathtruncatemacro{\nexty}{\y+1}
                
                \edef\drawedge{1}
                \ifnum\x=3 \ifnum\y=2 \edef\drawedge{0} \fi \fi
                
                \ifnum\drawedge=1
                  \draw (qb2-\x-\y-dmnd.north) -- (qb2-\x-\nexty-dmnd.south);
                \fi
              \fi
              
            }
          }
          
          \foreach \x in {1,...,\scxdist} {
            \draw[dashed] (qb2-\x-1-dmnd.south) -- +(0, -0.6);
            \draw[dashed] (qb2-\x-\scydist-dmnd.north) -- +(0, 0.6);
          }
          \foreach \y in {1,...,\scydist} {
            \draw[dashed] (qb2-1-\y-dmnd.west) -- +(-0.6, 0);
            \draw[dashed] (qb2-\scxdist-\y-dmnd.east) -- +(0.6, 0);
          }
        \end{scope}
      \end{pgfonlayer}

      \node[plabel] at ($(qb2-2-2-dmnd.center)!.5!(qb2-4-3-dmnd.center)$) {\(B_1 B_2\)};
    \end{scope}

  \end{tikzpicture}